\documentclass[showpacs,preprintnumbers,amsmath,amssymb,twocolumn,aps,pre]{revtex4-1}
\usepackage{color}
\usepackage{amsmath}
\usepackage{mathrsfs}
\usepackage{graphicx}
\usepackage{subfig}
\usepackage{epsfig}

\begin{document}
\title{Localized structures in dispersive and doubly resonant optical parametric oscillators}
\author{P. Parra-Rivas$^{1,2}$, L. Gelens$^{2}$, and F. Leo$^1$}

\affiliation{
$^1$OPERA-photonics, Université libre de Bruxelles, 50 Avenue F. D. Roosevelt, CP 194/5, B-1050 Bruxelles, Belgium\\
$^2$Laboratory of Dynamics in Biological Systems, KU Leuven Department of Cellular and Molecular Medicine, University of Leuven, B-3000 Leuven, Belgium\\}
\date{\today}


\begin{abstract}
We study temporally localized structures in doubly resonant degenerate optical parametric oscillators in the absence of temporal walk-off. We focus on states formed through the locking of domain walls between the zero and a non-zero continuous wave solution. We show that these states undergo collapsed snaking and we characterize their dynamics in the parameter space. 
\end{abstract}
\maketitle

\section{Introduction}
Localized structures (LSs) can be understood as domains of a finite size enclosed by stationary interfaces, and therefore their origin is usually related with the presence of bistability between two steady states \cite{coullet_nature_1987,coullet_localized_2002,tlidi_localized_1994}. In nature, they may appear in many different contexts ranging from vegetation patches in semi-arid regions or in sea grass ecosystems, to localized spots of light in driven nonlinear optical cavities \cite{fernandez-oto_c._strong_2014,ruiz-reynes_fairy_2017,willebrand_experimental_1993,umbanhowar_localized_1996,taranenko_patterns_2000,ramazza_localized_2000,barland_cavity_2002,leo_temporal_2010}. 

LSs are a particular type of so-called {\it dissipative structures} that emerge in systems far from the thermodynamic equilibrium due to a self-organization process \cite{nicolis_self-organization_1977}. Dissipative LSs arise due to a double balance between spatial coupling and nonlinearity on the one hand, and gain and dissipation on the other hand \cite{akhmediev_dissipative_2005}. Spatial coupling appears for example through the dispersion and/or dispersion of the light in optical systems, and is associated with diffusion processes in chemistry, biology and ecology \cite{murray_mathematical_2002,clerc_patterns_2005}. 

In optics, dissipative LSs have been widely studied in the context of externally driven diffractive nonlinear cavities with either cubic $\chi^{(3)}$ (Kerr) \cite{scroggie_pattern_1994,firth_two-dimensional_1996} or quadratic $\chi^{(2)}$ nonlinear media \cite{etrich_solitary_1997,staliunas_localized_1997,staliunas_spatial-localized_1998, longhi_localized_1997,oppo_domain_1999,oppo_characterization_2001,staliunas_transverse_2003}. In these cavities, LSs form in the plane transverse to the propagation direction, and they are commonly known as spatial cavity solitons. LSs have been also studied in wave-guided dispersive Kerr cavities, where LSs correspond to temporal pulses arising along the propagation direction, and they are one-dimensional \cite{leo_temporal_2010,chembo_spatiotemporal_2013,leo_dynamics_2013,herr_temporal_2014}. Temporal LSs have been considered as the basis for all-optical buffering \cite{leo_temporal_2010}, and in the last decade, also for the generation of broadband frequency combs in microresonators \cite{delhaye_optical_2007,kippenberg_microresonator-based_2011,pasquazi_micro-combs:_2018}. 
 
Recently, it has been shown that dispersive cavities with quadratic nonlinearities
may provide an alternative to Kerr cavities for the generation of frequency combs \cite{leo_walk-off-induced_2016,leo_frequency-comb_2016, mosca_frequency_2017,mosca_modulation_2018,hansson_quadratic_2018}. In contrast to Kerr combs, quadratic ones may operate with decreased pump power and can reach spectral regions that were not accessible before. Therefore, understanding the formation of temporal LSs is important in this context.

\begin{figure}[t]
	\centering
	\includegraphics[scale=0.45]{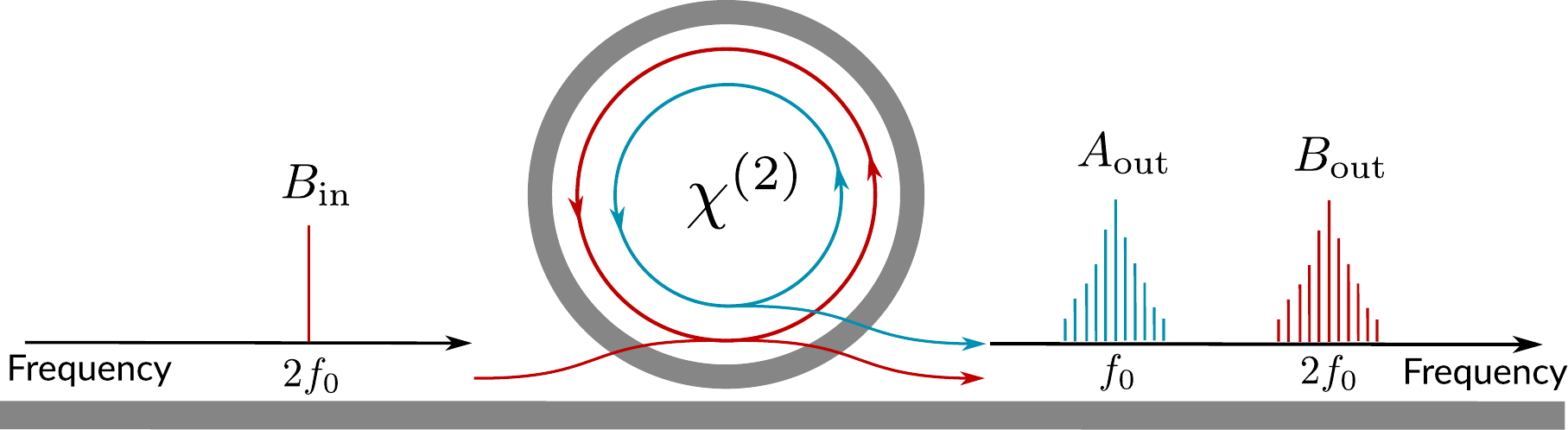}
	\caption{(Color online) Schematic example of a doubly resonant DOPO. A ring resonator with a $\chi^{(2)}$ nonlinearity is driven by a CW field $B_{in}$ at frequency $2f_0$. The quadratic interaction gives rise to a field $A$ with carrier frequency $f_0$ that resonates together with $B$, and therefore to two frequency spectra around $2f_0$ and $f_0$, respectively.}
	\label{sketch_DOPO}
\end{figure}

In this work we study the formation of LSs through the locking of domain walls (DWs) in a $\chi^{(2)}$-dispersive cavity matched for degenerate optical parametric oscillations (DOPO). A schematic example of such type of cavity is shown in Fig.~\ref{sketch_DOPO}. The cavity is externally driven by a pump field $B_{in}$ at frequency $2f_0$, and a field $A$ is generated at frequency $f_0$ through parametric down conversion. We consider a doubly resonant configuration such that both fields $A$ and $B$ resonate together in the cavity.
In such systems, continuous-wave (CW) states may coexist for the same values of a control parameter (bistability), and DWs connecting them can eventually form. DWs, also known as wave fronts or switching waves, exhibit a particle-like behavior in such a way that they can interact and lock, thus forming LSs of different extensions \cite{coullet_nature_1987,coullet_localized_2002}.

DWs have been previously studied in the context of diffractive DOPOs  \cite{trillo_stable_1997,oppo_characterization_2001}, and the formation of LSs through their locking has been analyzed in detail for both singly and doubly resonant configurations \cite{oppo_domain_1999, oppo_characterization_2001}. Recently, the formation of LSs has also been studied in dispersive DOPOs and in the presence of temporal walk-off \cite{parra-rivas_frequency_2019}. 

In all these studies, DWs and LSs form between CW states that have the same amplitude and are equally stable. As such they are also called equivalent CW solutions. However, in DOPOs, bistability between non-equivalent CW states is also present, and DWs and LSs may arise as well. Nonetheless, as far as we know, the formation of this type of LSs has not been analyzed in detail, neither in diffractive nor in dispersive cavities. Hence, in this paper we elucidate the formation, dynamics and bifurcation structure of the last type of LSs (hereafter type-I) and their connection with the former LSs (type-II). In this work we neglect the effect of the temporal walk-off.

The manuscript is organized as follows. In Sec.~\ref{sec:2} we introduce the mean-field model describing doubly resonant dispersive DOPOs and derive a single model with a nonlocal nonlinearity. In Section~\ref{sec:3} we present the stationary problem, analyze the CW solutions and their linear stability, and introduce the locking of DWs as the mechanism behind the formation of LSs. Later, in Sec.~\ref{sec:4} we calculate, applying multi-scale perturbation methods, a weakly nonlinear pulse-like solution about the trivial CW state. From Secs.~\ref{sec:5} to \ref{sec:7}, we then study the bifurcation structure of the different types of LSs formed through the locking of DWs, and how this structure is modified when varying the control parameters of the system. Finally, in Sec.~\ref{sec:8}, we discuss the main results of the paper.

\section{Mean-field models}\label{sec:2}
In this section we introduce the mean-field model for a dispersive DOPO in a doubly resonant configuration and we derive a nonlinear nonlocal model that will be used in the remainder of this work.

 
Assuming that the resonator exhibits high finesse, that both fields do not vary significantly over a single round-trip (i.e., the combined effects of nonlinearity and dispersion are weak), and following Refs.~\cite{haelterman_dissipative_1992,leo_frequency-comb_2016}, the dynamics of a DOPO can be described by a mean-field model for the slowly varying envelopes of the signal electric field $A$ centered at frequency $\omega_0$ and, the pump field $B$ centered at the frequency $2\omega_0$, as already shown in Ref.~\cite{parra-rivas_frequency_2019}. The normalized mean-field model reads:
\begin{subequations}\label{MF}
\begin{equation}\label{MF1}
\partial_t A=-(1+i\Delta_1)A-i\eta_1\partial_{x}^2A+i B\bar{A}
\end{equation}
\begin{equation}\label{MF2}
\partial_t B=-(\alpha+i\Delta_2)B-\left(d\partial_{x}+i\eta_2\partial_{x}^2\right)B+i A^2 +S.
\end{equation}
\end{subequations}
In the current formulation, $t$ corresponds to the normalized slow time describing the evolution of fields after every round-trip at a fixed position in the cavity, and $x$ is the normalized fast time \cite{parra-rivas_frequency_2019}. The parameter $\alpha$ is the ratio of the round-trip losses $\alpha_{1,2}$ associated with the propagation of the signal and pump fields, $\Delta_{1,2}$ are the normalized cavity phase detunings, $\eta_{1,2}$ and the group velocity dispersion (GVD) parameters of $A$ and $B$, $d$ is the normalized rate of temporal walk-off or wavevector mismatch related with the difference of group velocities between both fields, and $S$ is the driven field amplitude or pump at frequency $2\omega_0$. With the normalization used here $\eta_1=+1$($-1$) denotes normal (anomalous) GVD, and $\eta_2$ can take any value positive or negative.

The system of equations~(\ref{MF}) are formally equivalent to those describing diffractive spatial cavities \cite{oppo_formation_1994,zambrini_convection-induced_2005}. In that context, $\eta_1\approx 2\eta_2$ with $\eta_j>0$ are the diffraction parameters, $x$ represents a transverse spatial dimension, and $\partial_x^2$ applied to either $A$ and $B$ the beam diffraction. 

In contrast to spatial cavities, where the walk-off is normally negligible, in dispersive cavities it is very large and should be taken into consideration. The walk-off imposes severe restrictions on the efficiency of optical parametric amplification and often prevents the formation of LSs. Hence, it would be desirable to suppress it. This can be done by dispersion engineering as already shown in \cite{hansson_quadratic_2018}. Thus, in the following we will consider $d=0$. The effects of the walk-off on the stability and dynamics of LSs is beyond the scope of the present paper, and will be presented elsewhere. Furthermore, we will consider perfect phase-matching, what in wave-guided systems, as the one discussed here, implies $\Delta_2=2\Delta_1$.

The numerical exploration of the dynamics of Eqs.~(\ref{MF}) for a large range of parameters suggests that the $B$ field varies slowly in $t$. Thus, assuming 
that $\partial_t B\approx 0$, and following Refs.~\cite{nikolov_quadratic_2003,leo_frequency-comb_2016,parra-rivas_frequency_2019}, we can further simplify Eqs.~(\ref{MF}) to a single mean-field model for $A$ [see Appendix~A] with a nonlocal nonlinearity: 
 \begin{equation}\label{nl_GL}
\partial_t {\mathsf A}=-(1+i\Delta_1)\mathsf{A}-i\eta_1\partial_{x}^2\mathsf{A}-\bar{\mathsf{A}}(\mathsf{A}^2\otimes\mathsf{J}) +\rho\bar{\mathsf{A}},
\end{equation}
where $\otimes$ denotes convolution with the nonlocal kernel 
\begin{equation}
\mathsf{J}(x)=\frac{1+\tilde{\Delta}_2^{2}}{2\pi}\int_{-\infty}^{\infty}\frac{e^{-ikx}dk}{1+i(\tilde{\Delta}_2-\tilde{\eta}_2 k^2)},
\end{equation}
with 
$\tilde{\Delta}_2=\Delta_2/\alpha$, $\tilde{\eta}_2=\eta_2/\alpha$, although in the following we drop $(\tilde{\cdot})$.

The normalized field reads
\begin{equation}
\mathsf{A}=\frac{Ae^{-i\psi}}{\sqrt{\alpha(1+\Delta_2^2)}}
\end{equation}
with 
\begin{equation}
\psi=\pi/4+{\rm atan}(-\Delta_2)/2,
\end{equation}
and the normalized pump amplitude
\begin{equation}
\rho=\frac{S}{\alpha\sqrt{1+\Delta_2^2}}.
\end{equation}
Equation~(\ref{nl_GL}) is a kind of parametrically forced Ginzburg-Landau (PFGL) equation \cite{burke_classification_2008} with a long range coupling in $x$ introduced by the nonlocal nonlinearity $\mathsf{A}^2\otimes \mathsf{J}$. 
In this framework the interaction between $A$ and $B$ is equivalent to the propagation of $A$ in a medium with a nonlocal nonlinearity leading to an effective third order nonlinearity.

With this approximation, the $B$ field is dynamically slaved to $\mathsf{A}$, and explicitly given by
\begin{equation}\label{B_slaved}
B=(-\mathsf{A}^2\otimes\mathsf{J}+\rho)e^{i { \rm atan}(-\Delta_2)}.
\end{equation}
Models with a similar type of nonlocal response have already been considered in single-pass problems \cite{nikolov_quadratic_2003} and in quadratic dispersive cavities \cite{leo_walk-off-induced_2016,leo_frequency-comb_2016,mosca_frequency_2017,mosca_modulation_2018}. In particular Eq.~(\ref{nl_GL}) is 
formally equivalent to the mean field model derived in \cite{mosca_frequency_2017,mosca_modulation_2018} for the description of a singly resonant DOPO (with a different response function).

In all these cases the nonlocal response in Eq.~(\ref{nl_GL}) depends on $\mathsf{A}^2$, in contrast to other nonlocl models describing Raman \cite{lin_raman_2006,
chembo_spatiotemporal_2015}, diffusion \cite{krolikowski_modulational_2001,
suter_stabilization_1993} or thermal \cite{krolikowski_modulational_2001,firth_proposed_2007,
gordon_longtransient_1965} effects, where the nonlocal response depends on the intensity $|\mathsf{A}|^2$.

The models (\ref{MF}) and ($\ref{nl_GL}$) are equivalent when studying stationary states, such as LSs. Unless stated otherwise, here we focus on the study of Eq.~(\ref{nl_GL}). 


In terms of the real and imaginary part of $\mathsf{A}=U+iV$ Eq.~(\ref{nl_GL}) yields the system
\begin{equation}\label{nl_GL_real}
\partial_t\left[\begin{array}{c}
U \\ V
\end{array}\right]=\left(\mathcal{L}+\mathcal{N}\right)\left[\begin{array}{c}
U \\ V
\end{array}\right], 
\end{equation}
with $\mathcal{L}$ and $\mathcal{N}$ being the linear and nonlinear operators defined by
\begin{equation}
\mathcal{L}=\left[\begin{array}{cc}
\rho-1 & \Delta_1+\eta_1\partial_{xx} \\ -\Delta_1-\eta_1\partial_{xx} &-(\rho+1)
\end{array}
\right]
\end{equation}
and 
\begin{equation}
\mathcal{N}=-\left[\begin{array}{cc}
 \mathcal{N}^a& \mathcal{N}^b \\
\mathcal{N}^b & -\mathcal{N}^a
\end{array}\right],
\end{equation}
 with coefficients 
\begin{subequations}
	\begin{equation}
	\mathcal{N}^a=U^2\otimes \mathsf{J}_R-V^2\otimes \mathsf{J}_R-2UV\otimes \mathsf{J}_I,
	\end{equation}
	\begin{equation}
	\mathcal{N}^b=U^2\otimes \mathsf{J}_I-V^2\otimes \mathsf{J}_I+2UV\otimes \mathsf{J}_R,
	\end{equation}
\end{subequations}
where $\mathsf{J}_R$ and $\mathsf{J}_I$ correspond to the real and imaginary parts of the Kernel $\mathsf{J}$ [see Appendix A].
In the following we focus on the normal GVD regime ($\eta_1=+1$), and choose $\alpha=1$.

\section{Stationary solutions}\label{sec:3}
In this work we focus on the study of stationary states. In the current mean-field formulation these states satisfy $(\partial_t A,\partial_t B)=(0,0)$. They are thus solutions of the integro-differential equation:
\begin{equation}\label{NL_sta}
-i\eta_1\mathsf{A}_{xx}-(1+i\Delta_1)\mathsf{A}-\bar{\mathsf{A}} (\mathsf{A}^2\otimes \mathsf{J})+\rho\bar{\mathsf{A}}=0,
\end{equation}
or, equivalently, stationary states are solutions of
\begin{equation}\label{NL_sta2}
\left(\mathcal{L}+\mathcal{N}\right)\left[\begin{array}{c}
U \\ V
\end{array}\right]=\left[\begin{array}{c}
0 \\ 0
\end{array}\right]. 
\end{equation}
Stationary states can be of different nature such as homogeneous CW states \cite{lugiato_bistability_1988}, periodic patterns \cite{oppo_formation_1994,de_valcarcel_transverse_1996}, or DWs and LSs \cite{trillo_stable_1997,oppo_characterization_2001}.
Notice that equations~(\ref{nl_GL}) and (\ref{NL_sta}) are invariant under the transformations $x\rightarrow-x$, and $\mathsf{A}\rightarrow -\mathsf{A}$. The first symmetry means that any stationary solution is left/right symmetric (i.e. has a reflection symmetry), and according to the second symmetry, if $\mathsf{A}_s$ is a solution, so is $-\mathsf{A}_s$.

As stated before, in this work we focus on the study of LSs formed through the locking of DWs connecting different CWs. Hence, in this section we introduce the CW solutions of Eq.~(\ref{NL_sta}), analyze their linear stability, and study the formation of LSs.

\subsection{Continuous wave solutions}
The CW states of this system were first studied in Ref.~\cite{lugiato_bistability_1988} in the context of diffractive cavities. Here we review some of the results of that study in terms of the nonlinear nonlocal model (\ref{nl_GL}). In this framework the CWs correspond to the homogeneous steady state solutions of Eq.~(\ref{nl_GL}), which satisfy the algebraic equation:
\begin{equation}\label{HSS_sol}
-(1+i\Delta_1)\mathsf{A}_s-(1-i\Delta_2)\mathsf{A}_s|\mathsf{A}_s|^2+\rho \bar{\mathsf{A}}_s=0.
\end{equation} 
Writing $\mathsf{A}_s=|\mathsf{A}_s|e^{i\phi}$, Eq.~(\ref{HSS_sol}) becomes 
\begin{equation}
\left[-(1+i\Delta_1)-(1-i\Delta_2)|\mathsf{A}_s|^2+\rho e^{-2i\phi}\right]|\mathsf{A}_s|=0,
\end{equation}
\begin{figure}[t]
	\centering
	\includegraphics[scale=1]{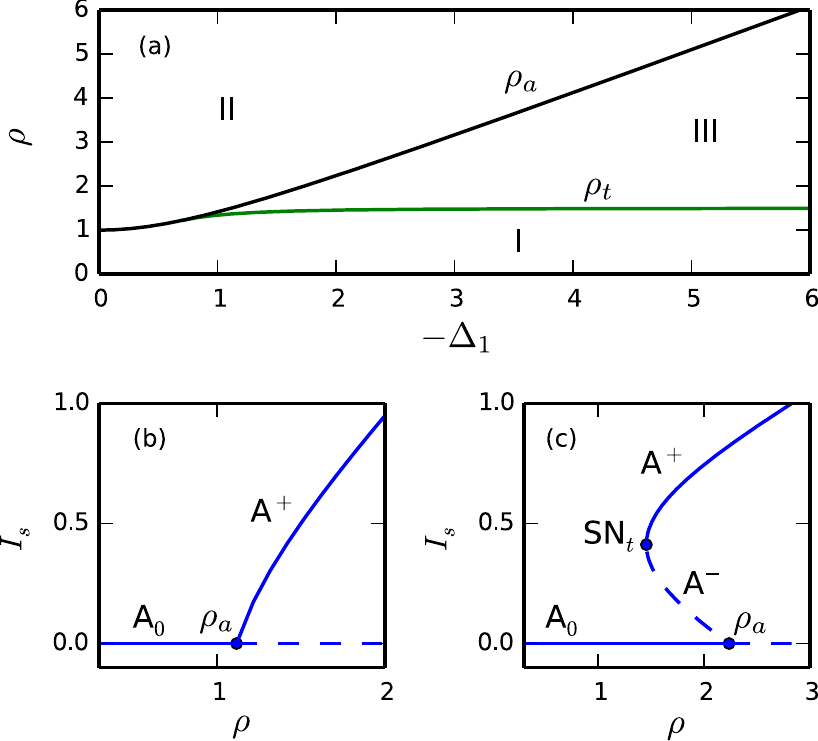}
	\caption{(Color online) In (a) the phase diagram in the $(\Delta_1,\rho)-$parameter space showing the principal bifurcation lines of the CW solutions: the pitchfork bifurcation $\rho_a$ (black line), and the fold or turning line $\rho_t$ corresponding to SN$_t$ (green line). (b) shows the CW solutions for $\Delta_1=-0.5$, and (c) those for $\Delta_1=-2$. The linear stability respect to homogeneous perturbations is shown using solid (dashed) lines for stable (unstable) states. The different regions are labeled by I-III and their description is given in the main text.}
	\label{F_sol_hom}
\end{figure} 
which yields three solutions: the trivial state $\mathsf{A}_s=\mathsf{A}_0=0$, and the two non-trivial states $\mathsf{A}^\pm=|\mathsf{A}^\pm|e^{i\phi^\pm}$, with 
\begin{equation}\label{hom}
|\mathsf{A}^{\pm}|^2=\frac{(\Delta_1\Delta_2-1)\pm\sqrt{(1+\Delta_2^2)\rho^2-(\Delta_2+\Delta_1)^2}}{1+\Delta_2^2},
\end{equation}
and phase 
\begin{equation}
	\phi^\pm={\rm acos}\left[(|\mathsf{A}^{\pm}|^2+1)/\rho\right]/2.
\end{equation}
If $\Delta_2\Delta_1>1$, only the $\mathsf{A}^+$ branch exists, and bifurcates super-critically from a pitchfork bifurcation \cite{haragus_local_2011} occurring at pump strength 
\begin{equation}
\rho_a=\sqrt{1+\Delta_1^2}. 
\end{equation}
The pitchfork bifurcation defines a line in the phase diagram in the $(\Delta_1,\rho)-$parameter space plotted in Fig.~\ref{F_sol_hom}(a) [see solid black line].
An example of the HSS bifurcation diagram in the super-critical regime is shown in Fig.~\ref{F_sol_hom}(b) for $\Delta_1=-0.5$. In contrast, for  $\Delta_2\Delta_1<1$, $\mathsf{A}^-$  arises sub-critically  as shown in Fig.~\ref{F_sol_hom}(c) for $\Delta_1=-2$, and undergoes a fold or turning point \cite{haragus_local_2011} at
\begin{equation}
\rho_t=\frac{\Delta_2+\Delta_1}{\sqrt{1+\Delta_2^2}},
\end{equation}
where it merges with $\mathsf{A}^+$. This line is plotted in green in Fig.~\ref{F_sol_hom}(a).  The transition between these two regimes occurs at a degenerate point at exactly $\Delta_2\Delta_1=1$, or equivalently $\Delta_1=1/\sqrt{2}$. 

We can therefore identify three main regions in the phase diagram of Fig.~\ref{F_sol_hom}(a):
\begin{itemize}
\item Region I: Only $\mathsf{A}_0$ exists and is stable. This region is spanned by the parameter region below $\rho_a$ for $\Delta_1<1/\sqrt{2}$, and $\rho<\rho_t$ for $\Delta_1>1/\sqrt{2}$.
\item Region II: The non-trivial solution $\mathsf{A}^+$ coexists with $\mathsf{A}_0$ that is now unstable. This region is spanned by $\rho>\rho_a$.
\item Region III: Solutions $\mathsf{A}_0$, $\mathsf{A}^-$ and $\mathsf{A}^+$ coexist, where $\mathsf{A}_0$ and $\mathsf{A}^+$ are both stable. This region is spanned by the values of $\rho$ such that $\rho_t<\rho<\rho_a$.
\end{itemize}

\subsection{Linear stability analysis of the continuous wave solutions}
Here we perform the linear stability analysis of the CW solutions in the presence of dispersion. Dispersion can cause the emergence of pattern forming instabilities, such as the Turing or modulational instability (MI) \cite{turing_alan_mathison_chemical_1952}. 
In the absence of dispersion, it is known that $A_0$ can undergo a Hopf instability leading to self oscillations, period doubling and chaos \cite{lugiato_bistability_1988}. Later the analysis was extended to include the effect of diffraction in the context of spatial cavities \cite{longhi_localized_1997}, and the spatio-temporal dynamics arising from the interaction of the Turing and Hopf modes was examined in detail in Refs.~\cite{tlidi_spatiotemporal_1997,tlidi_robust_1998}.  In this work we focus on the bistable regime ($\Delta_1\Delta_2<1$),  where self-pulsing of the CW states does not exist. In this context the linear stability of the CW can be analyzed by using the model (\ref{nl_GL}) instead of Eqs.~(\ref{MF}).
 
To perform this analysis we insert in Eq.~(\ref{nl_GL}) the ansatz 
\begin{equation}
	\mathsf{A}(t,x)=\mathsf{A}_s+\epsilon \zeta e^{\sigma t+ik x}+c.c.,
\end{equation}
describing a small modulation about the CW $\mathsf{A}_s$, 
where $\sigma$ is the growth rate of the perturbation, and $\zeta$ the eigenvector associated with the linearization of Eq.~(\ref{nl_GL}) at order $\epsilon$. The linear problem has modulated solutions if the growth rate satisfies
\begin{equation}\label{dispersion_relation}
\sigma^2+a_1\sigma +a_0=0,
\end{equation}  
where
\begin{subequations}
\begin{equation}
a_1=2(1+2I_s\mathcal{F}[\mathsf{J}_R]),
\end{equation}
\begin{equation}
a_0=c_2I_s^2+c_1I_s+c_0,
\end{equation}
\end{subequations}
and
\begin{subequations}
\begin{equation}
c_2=4\left(\mathcal{F}[\mathsf{J}_R]-(\eta_1k^2-\Delta_1)\mathcal{F}[\mathsf{J}_I]\right)
\end{equation}
\begin{equation}
c_1=4(\mathcal{F}[\mathsf{J}_R]^2+\mathcal{F}[\mathsf{J}_I]^2),
\end{equation}
\begin{equation}
c_0=\eta_1^2k^4-2\eta_1\Delta_1k^2.
\end{equation}
\end{subequations}
Here $\mathcal{F}$ denotes the Fourier transform as defined in Appendix~A.
The CW solutions $\mathsf{A}_0$ and $\mathsf{A}^{\pm}$ are linearly stable to perturbations with a given $k$ if Re$[\sigma(k)]<0$, and unstable otherwise. When $k=0$ we recover the homogeneous stability analysis performed in Ref.~\cite{lugiato_bistability_1988}, however when $k$ is allowed to vary the system can undergo a MI and periodic patterns may appear.

Through a linear stability analysis of the trivial solutions $\mathsf{A}_s=\mathsf{A}_0$, we obtain that $\mathsf{A}_0$ undergoes a MI at  
\begin{equation}
\rho=\rho_c\equiv 1,
\end{equation}
where patterns with a characteristic wavenumber
\begin{equation}
k_c=\sqrt{\eta_1\Delta_1},
\end{equation}
arise, provided that $\eta_1\Delta_1>0$. Two situations can be distinguished depending on the sign of the product $\eta_1\Delta_1$. When $\eta_1=1$ (normal GVD regime), $\mathsf{A}_0$ undergoes a MI if $\Delta_1>0$. In contrast, when $\eta_1=-1$ (anomalous regime), the MI occurs if $\Delta_1<0$. Notice that the stability of the trivial state does not depend on $\eta_2$.
  
The linear stability analysis of the non-trivial CW states $\mathsf{A}^{\pm}$ is cumbersome and an exact analytical expression of the MI threshold and critical wavenumber do not exist \cite{longhi_localized_1997}. Nevertheless, we can analyze the stability of these states by means of the
marginal instability curve $I_s(k)$. This curve defines the band of unstable modes, and is composed by two branches $I^\pm_s(k)$ satisfying the quadratic equation obtained by setting $\sigma=0$ in Eq.~(\ref{dispersion_relation}):
\begin{equation}
c_2I_s^2+c_1I_s+c_0=0.
\end{equation}   
The CW state is unstable against a perturbation with a fixed $k$, if $I_s$ is inside the curve, i.e. $I_s(k)^-<I_s<I_s^+(k)$, and unstable otherwise. For $k\neq0$ the extrema $(k,I_s)=(k_c,I_c)$ of this curve define the MI. 

 \begin{figure}[t]
 	\centering
 	\includegraphics[scale=0.95]{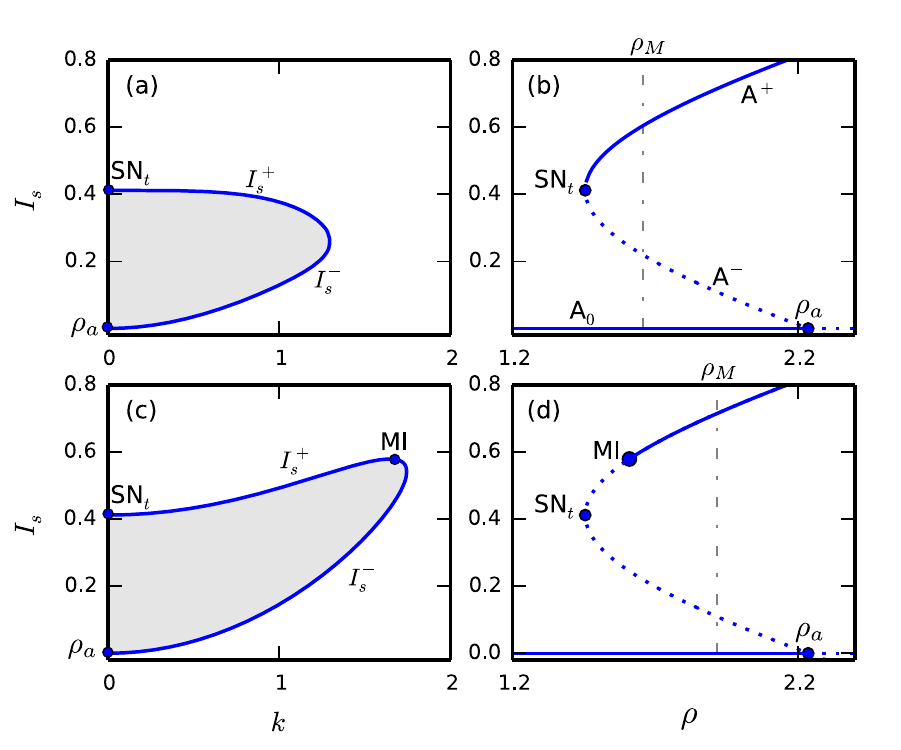}
 	\caption{(Color online) Panels (a)-(b) show the marginal instability curve and the bifurcation diagram associated with the CW solution for $(\Delta_1,\eta_2)=(-2,-0.8)$. Gray area in (a) shows the range of $I_s$ where the CW is unstable, and correspond to the dashed lines plotted in (b). The CW solution is stable outside this region as shown with solid lines in (a). Panels (c)-(d) show the same type of diagrams but for $(\Delta_1,\eta_2)=(-2,-0.05)$. The MI occurs at the maximum of this curve and is signaled with a blue dot in (d). The vertical gray dashed lines correspond to the Maxwell point $\rho_M$ of the system for such values of the parameters.}
 	\label{F_Marginal}
 \end{figure}

Figure~\ref{F_Marginal} (a) shows the marginal instability curve associated with the CW solution shown in panel (b) for $(\Delta_1,\eta_2)=(-2,-0.8)$.
The maximum of this curve occurs at $I_t$ for $k=0$, and therefore $\mathsf{A}^-$ is unstable from $\rho_a$ to SN$_t$ [see dotted line in Fig.~\ref{F_Marginal}(b)], while $\mathsf{A}^+$ is stable for any value of $k\neq0$ as shown in Fig.~\ref{F_Marginal}(b).

Decreasing the value of $|\eta_2|$ the maximum migrates from the fold SN$_t$ at $(k,I_s)=(0,I_t)$ to $(k,I_s)=(k_c,I_c)$ where a MI takes place. This is the situation shown in Fig.~\ref{F_Marginal}(c) for $\eta_2=-0.05$. In this case $\mathsf{A}^-$ remains unstable, and $\mathsf{A}^+$ is stable above the MI, i.e. for $I_s>I_c$, and unstable otherwise [see solid and dotted lines in Fig.~\ref{F_Marginal}(d)]. 

The MI defines a manifold $\rho_c=\rho(I_s(k_c),\eta_2,\Delta_1)$ according to which region III can be subdivided as follows:
\begin{itemize}
\item III$_a$: $\mathsf{A}^+$ is unstable in response to non-homogeneous perturbations (i.e. $k\neq0$). This region spans the parameter space $\rho_t<\rho<\rho_c$.
\item III$_b$: $\mathsf{A}^+$ is stable in response to non-homogeneous perturbations. This region spans the parameter region $\rho>\rho_c$. 
\end{itemize}

\subsection{Formation of localized states through domain wall locking}

\begin{figure*}[t]
	\centering
	\includegraphics[scale=0.65]{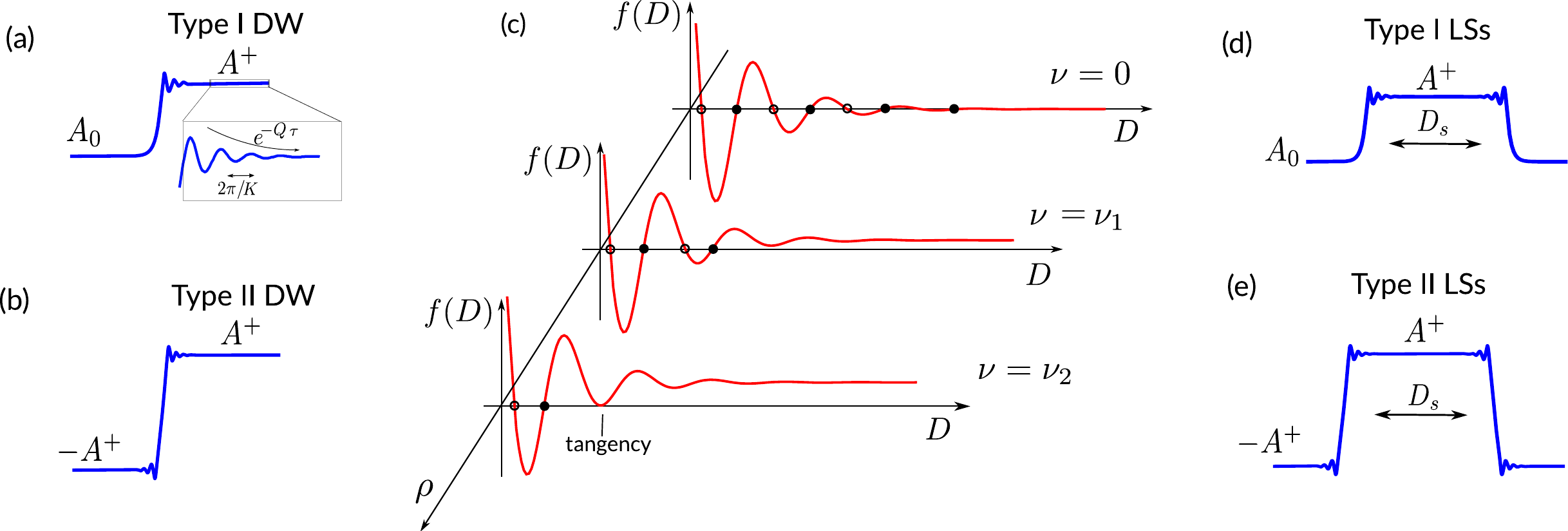}
	\caption{(Color online) (a,b)  The real component of a DW of type-I ($\mathsf{A}_0\rightarrow \mathsf{A}^+$) (a) and a DW of type-II ($-\mathsf{A}^+\rightarrow \mathsf{A}^+$) (b). (c) Sketch of the oscillatory interaction defined by Eq.~(\ref{DW_interact_I}) at the Maxwell point ($\nu=0$) and two locations away from the Maxwell point (i.e. $\nu=\nu_1$ and $\nu_2$). The stable (unstable) separations $D_s$ are labeled using $\bullet$ ($\circ$); (d,e) Example of Type I (d) and Type II (e) LSs.}
	\label{DWs_LSs}
\end{figure*}

As shown in the previous sections, the CW solutions may coexist stably depending on the range of parameters. Therefore, in the presence of dispersion, DWs may arise connecting two different CWs. In this context two different types of DWs occur: 
\begin{itemize}
	\item Type-I: the connection occurs between $\mathsf{A}_0$ and $\mathsf{A}^+$ [see Fig.~\ref{DWs_LSs}(a), left]. They exist in region III$_b$.
	\item Type-II: the connection arises between two equivalent (equally stable) non-trivial states, i.e. $-\mathsf{A}^+$ and $\mathsf{A}^+$. They occur in regions II and III$_b$. [see Fig.~\ref{DWs_LSs}(b), left]
\end{itemize}
The tails of both type-I and type-II DWs around the CW-solution $\mathsf{A}^+$ [see close-up view in (a)] can be described asymptotically  by the ansatz $\mathsf{A}(x)=\mathsf{A}^{+}_s+\epsilon e^{\lambda x}+c.c.$, where the eigenvalues $\lambda$ satisfy the condition $\sigma(-i\lambda)=0$, and are therefore solutions of the polynomial  
\begin{equation}\label{charac_poli}
b_8\lambda^8+b_6\lambda^6+b_4\lambda^4+b_2\lambda^2+b_0=0,
\end{equation}
where the coefficients $b_m$ are functions of the parameters of the system.

Due to the reflection symmetry $x\rightarrow -x$, Eq.~(\ref{charac_poli}) is invariant under $\lambda\rightarrow-\lambda$, and $\lambda\rightarrow\bar\lambda$ \cite{champneys_homoclinic_1998}. Equation~(\ref{charac_poli}) cannot be solved analytically except in some particular conditions \cite{oppo_domain_1999}.
The tails can approach $\mathsf{A}^+$ either monotonically, or in a damped oscillatory fashion. The latter case is related with the existence of at least four complex eigenvalues $\lambda_{1,2,3,4}=\pm Q\pm iK$, those with the smallest real part $|Q|$. The oscillatory damped tails are described by
\begin{equation}
\mathsf{A}(x)=\mathsf{A}_s^++a{\rm cos}(Kx)e^{-Qx}.
\end{equation}
In contrast, when $K=0$ the oscillations disappear, and the DW approaches $\mathsf{A}^+$ monotonically. 
In what follows we separately describe the interaction of DWs and the formation of type-I and type-II LSs .

\subsubsection*{Type-I domain walls and localized structures}

The CW states $\mathsf{A}_0$ and $\mathsf{A}^{+}$ are non-equivalent, and type-I DWs move with a constant velocity that depends on the control parameters of the system.
In gradient systems, where an energy functional can be defined, the velocity is proportional to the energy difference between $\mathsf{A}_0$ and $\mathsf{A}^{+}$. In this context, the Maxwell point of the system is defined as the  parameter value where both CW states have the same energy, or equivalently, as the point where the velocity of the DWs becomes zero \cite{chomaz_absolute_1992}. 
Here, despite the
system not having gradient dynamics, we will still refer to such a point as the Maxwell point, and hereafter we label it as $\rho_M$. This point is marked using a dotted-dashed line in Figs.~\ref{F_Marginal}(b) and (d). In a range of parameters around $\rho_M$ two DWs 
with different polarity, let say a kink $\mathsf{A}_0\rightarrow \mathsf{A}^{+}$ and anti-kink $\mathsf{A}^+\rightarrow \mathsf{A}_0$, separated by a distance $D$ interact as described by 
\begin{equation}\label{DW_interact_I}
\partial_tD=\varrho {\rm cos}(KD)e^{-QD}+\nu\equiv f(D),
\end{equation} 
where $\nu\sim\rho-\rho_M$, measures the distance from the Maxwell point $\rho_M$, and $\varrho$ depends on the parameters of the system \cite{coullet_localized_2002}. 

When $K\neq0$, the oscillatory nature of the interaction leads to alternating regions of attraction and repulsion [see Fig.~\ref{DWs_LSs}(c)].  DWs lock at different stationary separations $D_s$ satisfying $f(D_s)=0$. At $\rho=\rho_M$ ($\nu=0$) [Fig.~\ref{DWs_LSs}(c), bottom], the width of the LSs ($D_s$) is quantized: $D_s^n=\frac{\pi}{2K}(2n+1)$, with $n=0,1,2,\dots$ \cite{coullet_nature_1987,coullet_localized_2002}. Figure~\ref{DWs_LSs}(d) shows an example of a LS of width $D_s$, corresponding to the stationary distances shown in Fig.~\ref{DWs_LSs}(c). The stable (unstable) separation distances are marked with $\bullet$ ($\circ$). We refer to these states as Type-I LSs.
When $\nu\neq0$ the red curve shifts upwards or downwards (depending on the sign of $\nu$), and as a result, the number of stationary intersections decreases as is shown in Fig.~\ref{DWs_LSs}(c) for $\nu=\nu_1$ and $\nu_2$. Hence, when moving away from the Maxwell point $\rho_M$, the widest LSs disappear first, but eventually even the single peak LS is lost.    
In Sec.~\ref{sec:5} we will see that the interaction described by Eq.~(\ref{DW_interact_I}) is responsible of the bifurcation structure that the previous LSs undergo.

When the tails are monotonic ($K=0$) a different phenomenon known as coarsening occurs where two DWs with different polarity attract each other until eventually they annihilate one another \cite{allen_microscopic_1979}.
\subsubsection*{Type-II domain walls and localized structures}
In regions II and III$_b$ the solutions $-\mathsf{A}^+$ and $\mathsf{A}^+$ coexist and are linearly stable, and hence type-II DWs connecting them may also arise. In this case these CWs are equivalent, and therefore, the DWs are stationary [see Fig.~\ref{DWs_LSs}(b)]. Here the interaction between kink $(-\mathsf{A}^+\rightarrow \mathsf{A}^{+})$ and anti-kink $(\mathsf{A}^+\rightarrow -\mathsf{A}^{+})$ is described by Eq.~(\ref{DW_interact_I}) by setting $\nu=0$ \cite{coullet_nature_1987} (see Fig.~\ref{DWs_LSs}(c)). The LSs resulting from this interaction are referred to as type-II LSs, and have been largely studied in the context of diffractive cavities \cite{oppo_domain_1999,oppo_characterization_2001}. DWs of this type may undergo non-equilibrium Ising-Bloch transition, where DWs start to drift \cite{coullet_breaking_1990}, and as result LSs may show very complex dynamics \cite{gomila_theory_2015}.
An example a such type of state is shown in Fig.~\ref{DWs_LSs}(e).

\section{Weakly-nonlinear solutions around the pitchfork bifurcation}\label{sec:4}
While the locking of DWs explains the formation of high amplitude LSs, it does not describe their origin from a bifurcation point of view. In this section we show that those structures are connected  with small amplitude states that arise from the Pitchfork bifurcation occurring at $\rho_a$. In order to do so we derive a stationary normal form for the pitchfork bifurcation by applying weakly nonlinear multi-scale analysis. We find two types of extended solutions that explain the origin of the structures discussed in Sec.~\ref{sec:3}. The solutions of this normal form have been studied in the context of the parametrically forced Ginzburg-Landau equation \cite{burke_classification_2008}. However, in our case, we have a long-range nonlocal coupling in $x$ in terms of the nonlocal nonlinearity $\mathsf{A}^2\otimes J$. In order to deal with this difficulty we follow the approach shown in Ref.~\cite{morgan_swifthohenberg_2014}.

Following~\cite{burke_classification_2008} we fix $\Delta_1$ and consider the asymptotic expansion of the fields $U$, and $V$ as a function of the expansion parameter $\epsilon$ defined by $\rho=\rho_a+\delta\epsilon^2$, where $\delta$ is the bifurcation parameter. Then the expansion reads
\begin{equation}\label{expand}
\left[\begin{array}{c}
U \\ V
\end{array}
\right]=\epsilon\left[\begin{array}{c}
u_1 \\ v_1
\end{array}
\right]+\epsilon^3\left[\begin{array}{c}
u_3 \\ v_3
\end{array}
\right]+\cdots.
\end{equation}
where we allow each of the terms in the previous expansion to depend just on the long scale $x_1\equiv\epsilon x$. Considering Eq.~(\ref{expand}) the linear operator expands as 
\begin{equation}
\mathcal{L}=\mathcal{L}_0+\epsilon^2\mathcal{L}_2,
\end{equation}
with
\begin{subequations}
	\begin{equation}
	\mathcal{L}_0=\left[\begin{array}{cc}
	\rho_a-1 & \Delta_1 \\ -\Delta_1 &-(\rho_a+1)
	\end{array}
	\right],
	\end{equation}
	and
	\begin{equation}
	\mathcal{L}_2=\left[\begin{array}{cc}
	\delta & \eta_1\partial^2_{x_1} \\ -\eta_1\partial^2_{x_1} & -\delta
	\end{array}
	\right].
	\end{equation}
\end{subequations}
Similarly the nonlinear operator becomes 
\begin{equation}
\mathcal{N}=\epsilon^2\mathcal{N}_2=-\left[\begin{array}{cc}
\mathcal{N}_2^a & \mathcal{N}_2^b\\
\mathcal{N}_2^b & -\mathcal{N}_2^a
\end{array}\right],
\end{equation}
with 
\begin{subequations}
	\begin{equation}
		\mathcal{N}_2^a=u_1^2\otimes \mathsf{J}_R-v_1^2\otimes \mathsf{J}_R-2u_1v_1\otimes \mathsf{J}_I  
	\end{equation}
	\begin{equation}
		\mathcal{N}_2^b= u_1^2\otimes \mathsf{J}_I-v_1^2\otimes \mathsf{J}_I+2u_1v_1\otimes \mathsf{J}_R.
	\end{equation}
\end{subequations}
The insertion of the previous expansions in the stationary equation (\ref{NL_sta2}) yields a hierarchy of equations for successive orders in $\epsilon$, which up to third order read: 
\begin{subequations}
\begin{equation}\mathcal{O}(\epsilon): 
\mathcal{L}_0\left[\begin{array}{c}
u_1 \\ v_1
\end{array}
\right]=\left[\begin{array}{c}
0 \\ 0
\end{array}
\right],		
	\end{equation}
	and 
	\begin{equation}
		 \mathcal{O}(\epsilon^3):\mathcal{L}_0\left[\begin{array}{c}
		 u_3 \\ v_3
		 \end{array}
		 \right]+(\mathcal{L}_2+\mathcal{N}_2)\left[\begin{array}{c}
		u_1\\v_1
		\end{array}
		\right]=\left[\begin{array}{c}
		0 \\ 0
		\end{array}
		\right]
	\end{equation}
\end{subequations}

At first order in $\epsilon$ the solvability condition provides,
\begin{equation}
\rho_a=\sqrt{\Delta_1^2+1},
\end{equation}
which confirms the position of the pitchfork bifurcation already calculated in Sec.~\ref{sec:3}.
The solutions at this order are of the form 
\begin{equation}
\left[\begin{array}{c}
u_1 \\ v_1
\end{array}\right]=\left[\begin{array}{c}
\xi \\ 1
\end{array}\right]a(x_1),
\end{equation}
where $\xi=\Delta_1/(1-\rho_a)$ and $a(x_1)$ is the real envelope amplitude to be determined at next order in the expansion.
\begin{figure}[t]
	\centering
	\includegraphics[scale=1]{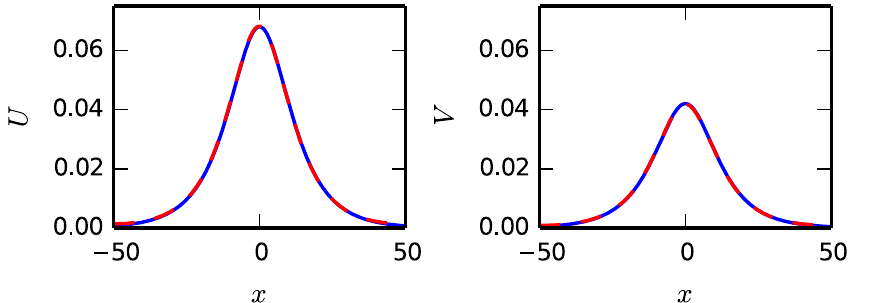}
	\caption{(Color online) Weakly nonlinear solution around the pitchfork bifurcation $\rho_a$. (a)-(b) show in blue the real and imaginary profiles of the weakly nonlinear state given by (\ref{bump}) for $\Delta_1=-2$ and $\rho-\rho_a=0.01$. Red dashed lines represent the numerical solutions of Eq.~(\ref{NL_sta}) at the same point. Both lines are indistinguishable.}
	\label{F_WNL_solutions}
\end{figure}
Applying the same procedure as in Ref.~\cite{morgan_swifthohenberg_2014} we show [see Appendix~B] that the 
the solvability condition at $\mathcal{O}(\epsilon^3)$ gives
the stationary normal form for the amplitude $a$:
\begin{equation}\label{normal_form}
C_1\partial^2_{x_1}a=\delta a+C_3a^3,
\end{equation}
with the coefficients
\begin{subequations}
	\begin{equation}
	C_1= \frac{-2\eta_1\xi}{1+\xi^2}
	\end{equation}
	and
	\begin{equation}
	C_3=1-\xi(\xi+2\Delta_2) 
	\end{equation}
\end{subequations}
This last equation admits the CW solutions $a=\sqrt{-\delta/C_3}$, or equivalently,
\begin{equation}
\left[\begin{array}{c}
U\\V
\end{array}
\right]=\left[\begin{array}{c}
\displaystyle\frac{\Delta_1}{1-\rho_a}\\1
\end{array}
\right]\sqrt{\frac{\rho-\rho_a}{-C_3}}+\cdots, 
\end{equation}
that confirms the result already obtained in Sec.~\ref{sec:3}: the CW bifurcates super-critically $(\rho>\rho_a)$ if $\Delta_2\Delta_1>1$, and sub-critically $(\rho<\rho_a)$ if $\Delta_2\Delta_1<1$.
 
In the super-critical regime ($\Delta_2\Delta_1>1$) the normal form (\ref{normal_form}) admits DW-like solutions of the form
\begin{equation}
a(x_1)=\sqrt{\frac{\delta}{-C_3}}{{\rm tanh}\left(\sqrt{\frac{\delta}{-2C_1}}x_1\right)},
\end{equation}
yielding a super-critical bifurcation to states of the form
\begin{equation}\label{small_DW}
\left[\begin{array}{c}
U\\V
\end{array}
\right]=\left[\begin{array}{c}
\displaystyle\frac{\Delta_1}{1-\rho_a}\\1
\end{array}
\right]\sqrt{\frac{\rho-\rho_a}{-C_3}}{{\rm tanh}\left(\sqrt{\frac{\rho-\rho_a}{-2C_1}}x\right)}+\cdots
\end{equation}
This analytical solution was first obtained in the context of diffractive cavities in Ref.~\cite{longhi_localized_1997}, where the normal form around$\rho_a$ was derived in terms of the full model~(\ref{MF}).

In contrast, for $\Delta_1\Delta_2<1$,
the normal form (\ref{normal_form}) admits solutions of the form  
\begin{equation}
	a(x_1)=\sqrt{\frac{-2\delta}{C_3}}{{\rm sech}\left(\sqrt{\frac{\delta}{C_1}}x_1\right)},
\end{equation}
which provides the sub-critical emergence of type-I LSs 
\begin{equation}\label{bump}
\left[\begin{array}{c}
U\\V
\end{array}
\right]=\left[\begin{array}{c}
\displaystyle\frac{\Delta_1}{1-\rho_a}\\1
\end{array}
\right]\sqrt{\frac{2(\rho_a-\rho)}{C_3}}{{\rm sech}\left(\sqrt{\frac{\rho_a-\rho}{-C_1}}x\right)}+\cdots,
\end{equation}
\begin{figure*}[t]
	\centering
	\includegraphics[scale=1.0]{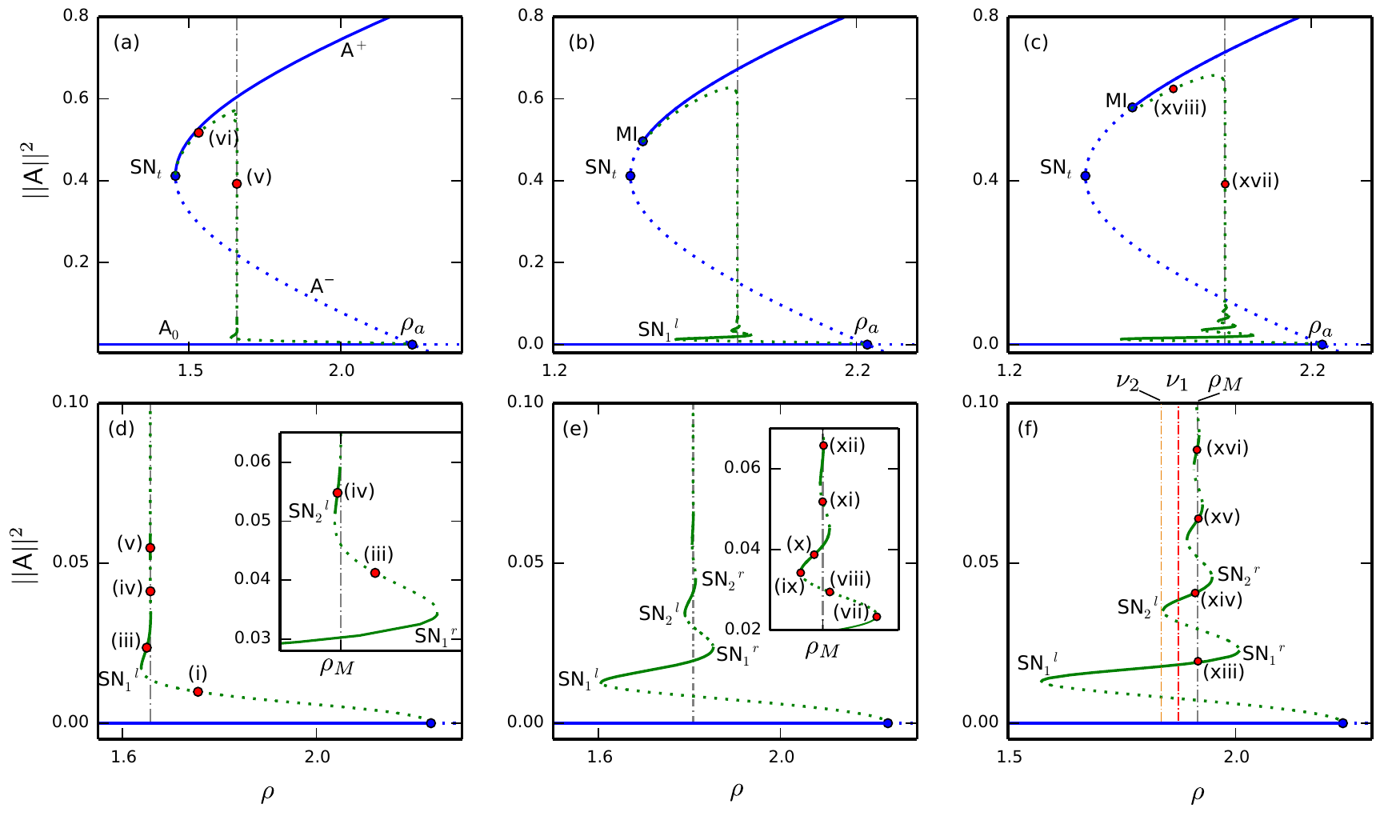}
	\includegraphics[scale=1.0]{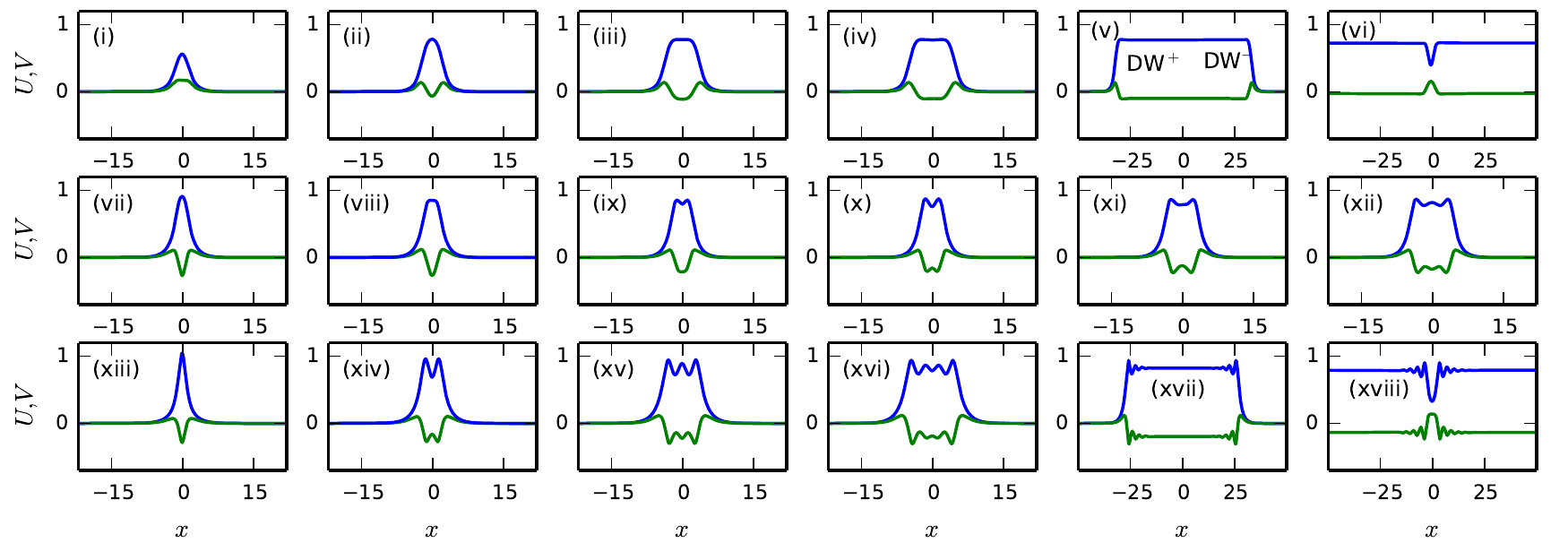}	
	\caption{(Color online) Bifurcation diagrams for LSs of type-I at $\Delta_1=-2$ and different values of $\eta_2$. In (a),(d) collapsed snaking for $\eta_2=-0.8$, in panels (b),(e) for $\eta_2=-0.2$, and (d),(f) correspond to $\eta_2=-0.05$. The panels (d), (e), and (f) are close-up views of the bottom parts of the bifurcation diagrams shown in (a), (b), and (c).  Solid (dashed) lines correspond to stable (unstable) solutions. The vertical gray point-dashed line stands for $\rho_M$, and red and orange vertical lines in panel (f) refer to Fig.~\ref{DWs_LSs}(c). The different SNs of the LSs are labeled through SN$_i^{l,r}$, and the red dots correspond to the LSs shown in the subpanels (i)-(xviii), where blue and green solid lines represent $U$ and $V$ respectively.}
	\label{bif_diagram_typeI}
\end{figure*}
These weakly-nonlinear solutions are only valid close to the pitchfork bifurcation at $\rho_a$. In the next section we show how these solutions are modified when entering the highly nonlinear regime as one of the control parameters of the system is changed. Notice that the weakly nonlinear solutions (\ref{small_DW}) and (\ref{bump}) are independent of the parameter $\eta_2$. This shows that in the weakly nonlinear regime the states studied here are not influenced by the presence of the long-range interaction in $x$. 

In the coming section we focus on the sub-critical regime and study the bifurcation structure of LSs of the form (\ref{bump}). To check the validity of our calculations, in Fig.~\ref{F_WNL_solutions} we have plotted the real and imaginary parts (blue line) of the weakly nonlinear state (\ref{bump}) together with the numerical solutions (dashed red line) obtained through a Newton-Raphson solver, showing excellent agreement. 

\section{Bifurcation structure of type-I localized states}\label{sec:5}
In this section we study the bifurcation structure of the type-I LSs. In Sec.~\ref{sec:4} we have derived a normal form equation around the pitchfork bifurcation occurring at $\rho_a$. Two stationary weakly nonlinear solutions are found corresponding to a small amplitude DW and bump [see Eq.~(\ref{small_DW}) and Eq.~(\ref{bump})] that arise in the super-critical and sub-critical regime, respectively. 

In what follows we focus on the sub-critical regime and, unless stated otherwise, fix $\Delta_1=-2$, and $\eta_2<0$. Weakly nonlinear solutions are only valid in a neighborhood of the bifurcation at $\rho_a$. However, applying numerical continuation techniques \cite{allgower_numerical_1990} we are able to track these solutions to parameter values away from the small amplitude bifurcation $\rho_a$, and therefore, to build bifurcation diagrams as those shown in Fig.~\ref{bif_diagram_typeI}. In these diagrams the $L^2$-norm $||\mathsf{A}||^2=\frac{1}{L}\int_{-L/2}^{L/2}|\mathsf{A}(x)|^2dx$ is plotted as a function of the pump intensity $\rho$ for different values of $\eta_2$.

Figures~\ref{bif_diagram_typeI}(a),(d) show the bifurcation diagram for $\eta_2=-0.8$, where panel (d) is a close-up view of the diagram shown in panel (a). The blue lines in Fig.~\ref{bif_diagram_typeI}(a) represent the CW solution, whose linear stability is shown using solid (dashed) lines for stable (unstable) solutions. The vertical gray line corresponds to the Maxwell point of the system $\rho_M$. At this point the velocity of the DWs connecting the trivial solution $\mathsf{A}_0$ with the non-trivial one $\mathsf{A}^+$ is zero, and around this point two DWs of different polarities can lock to each other and form LSs of type-I, as already discussed in Sec.~\ref{sec:3}. Close to $\rho_a$ the LS is well described by the small amplitude weakly nonlinear solution (\ref{bump}), and is initially unstable.

The stability of the $x-$dependent steady states is obtained from the analysis of the eigenspectrum of the linear operator associated with Eq.~(\ref{nl_GL}) evaluated at such steady state. This linear operator must be calculated numerically, and hence, it corresponds to the Jacobian matrix associated with the coupled algebraic equations that originate from discretizing Eq.~(\ref{nl_GL}). To confirm the validity of the stability results we have also performed such analysis using the full model (\ref{MF}). Indeed, for the type of states studied here, the stability analysis using both models agrees.

Decreasing $\rho$ the amplitude of the LSs increases [see profile (i)] until reaching the first fold of the diagram. This fold correspond to a saddle-node bifurcation that we label as SN$_1^l$ [see Fig.~\ref{bif_diagram_typeI}(d)]. Once SN$_1^l$ is passed the LS become stable. At this stage the LS corresponds to a high amplitude state as the one shown in panel (ii). Increasing $\rho$ further the amplitude of the LS grows, and it becomes unstable at a second saddle-node SN$_1^r$ [see inset]. 
At the same time a small dip is nucleated in the central position of the LS forming an almost flat plateau [see panel (iii)]. While increasing the norm the LS broadens and becomes stable one more time at SN$_2^l$ [see profile (iv)]. Proceeding up in the diagram (i.e. increasing $||\mathsf{A}||^2$) the process repeats, resulting in the broadening of the LSs as shown in panel (v). At this stage one can observe how the LS is formed by a pair of DWs connecting $\mathsf{A}_0$ with $\mathsf{A}^+$ of different polarities, namely DWs$^+$ and DWs$^-$.

In the course of this process the solution branches undergo a sequence of exponentially decaying oscillations in $\rho$ at the vicinity of the Maxwell point  $\rho_M\approx1.6578$ [see inset of Fig.~\ref{bif_diagram_typeI}(d)]. This type of bifurcation structure is known as {\it collapsed snaking} \cite{knobloch_homoclinic_2005,ma_defect-mediated_2010,burke_classification_2008}, and has been studied in detail in the context of Kerr cavities \cite{parra-rivas_dark_2016}. 

In periodic systems like ours the LS branch moves away from $\rho\approx\rho_M$ when the maximum amplitude starts to decrease below $\mathsf{A}^-$ and the solution turns into a dark LS sitting on $\mathsf{A}^+$ [see profile (vi) translated $L/2$]. This branch terminates at SN$_t$, where the amplitude of the LS becomes zero. In terms of spatial dynamics this point corresponds to a reversible Takens-Bogdanov bifurcation \cite{champneys_homoclinic_1998, haragus_local_2011}, and a weakly nonlinear solution of the form $\mathsf{A}-\mathsf{A}^+\sim a{\rm sech}^2(bx)$ can be obtained as already done in Refs.~\cite{parra-rivas_dark_2016,godey_bifurcation_2017,parra-rivas_bifurcation_2018}.


\begin{figure}[!t]
	\centering
	\includegraphics[scale=0.9]{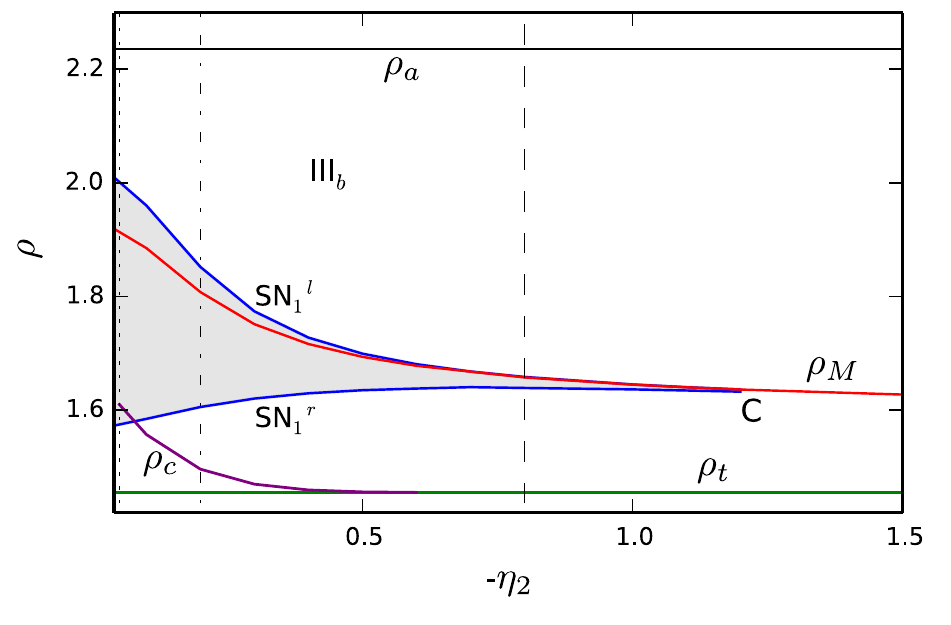}
	\caption{(Color online) Phase diagram in the $(\eta_2,\rho)-$parameter space for $\Delta_1=-4$. The gray area limited by SN$_1^r$ and SN$_1^l$ corresponds to the parameter region where LSs of type I exist. The red and purple lines correspond to the Maxwell point $\rho_M$ and the MI $\rho_c$ respectively. The horizontal bifurcation lines in black and green are the pitchfork bifurcation $\rho_a$ and the saddle-node bifurcation $\rho_t$ of the CW solution. The inset shows a close-up view about the cusp bifurcation (C) where SN$_1^r$ and SN$_1^l$ collide and disappear. The pointed, dashed-pointed, and dashed vertical lines correspond to the bifurcation diagrams shown in Fig.~\ref{bif_diagram_typeI} for $\Delta_1=-0.8,-0.2$, and $-0.05$.}
	\label{phase}
\end{figure}

The bifurcation diagrams shown in Fig.~\ref{bif_diagram_typeI}(a) and (d) correspond to a slice for constant $\eta_2=-0.8$ of the phase diagram shown in Fig.~\ref{phase} [dashed vertical line], where the main bifurcation lines are plotted in the $(\eta_2,\rho)-$parameter space for constant $\Delta_1=-2$. The saddle-node  and the pitchfork bifurcations of the CW $\rho_t$ and $\rho_a$ are plotted in black and green solid lines respectively. The Maxwell point $\rho_M$ is indicated with a red solid line, the MI $\rho_c$ is shown in purple, and the SN$_1^l$ and SN$_1^r$ are plotted in blue. The gray area in-between these lines is the region where type-I LSs exist. Increasing $\eta_2$ the different folds SN$_i^l$ and SN$_i ^r$ with $i=1,2,...$ approach one another and disappear in a sequence of cusp bifurcations. Here we only show the cusp that involves the collision of SN$_1^l$ and SN$_1^r$. 

When decreasing $|\eta_2|$, the situation is rather different. The MI instability, not present before, arises from SN$_t$ around $\eta_2\approx-0.6$ and separates from it when moving toward lower values of $\eta_2$, destabilizing the CW branch $A^+$. Figure~\ref{bif_diagram_typeI}(b),(e) shows the bifurcation diagram corresponding to this situation for $\eta_2=-0.2$. As in the previous case, a branch of LSs arises from the pitchfork bifurcation at $\rho_a$ and undergoes collapsed snaking. However, in this case, the Maxwell point, and the bifurcation diagram itself have shifted to higher values of $\rho$. Furthermore, while in Fig.~\ref{bif_diagram_typeI}(a)-(d) the solutions branches collapse rapidly to $\rho_M$ as increasing the $||\mathsf{A}||^2$, in panels (b)-(e) the collapse is much slower, and hence the solution branches of wider structures persist.

The profiles (vii-xii) show how the LSs are modified while passing through two consecutive folds [see Fig.~\ref{bif_diagram_typeI}(e)]. In (i) the the LS consist in a single bump. Soon after passing SN$^r_1$ the structure start to develop a central dip [see profile (vii)] that deepens as decreasing $\rho$ until reaching SN$^l_2$ [see (viii)] where it becomes stable. This process repeats: at every SN$^r_i$ a new dip is nucleated from the center of the LS which broadens as increasing $||\mathsf{A}||^2$ [see profiles (viii)-(xii)].

As before, the branch of LSs detaches from $\rho_M\approx1.8079$ when $||\mathsf{A}||^2\approx 0.6$, and persists until it meets with $\mathsf{A}^+$. Here, however, the merging occurs not at the SN$_t$, but at the MI at $\rho_c\approx1.4963$. 
Indeed close to the MI, one can show that a weakly nonlinear periodic pattern of wavelength $2\pi/k_c$ exist and arise sub-critically from $\rho_c$ together with a bump solutions of the form $\mathsf{A}-\mathsf{A}^+\sim a{\rm sech}(bx){\rm cos}(k_cx+\varphi)$, where $a$ and $b$ depend on the control parameters of the system, and $\varphi$ controls the phase of the pattern within the sech \cite{kozyreff_asymptotics_2006,parra-rivas_bifurcation_2018}. This structure is plotted in panel (xviii) of Fig.~\ref{bif_diagram_typeI} for $\eta_2=-0.05$. 
These type of LSs may undergo {\it homoclinic snaking} \cite{woods_heteroclinic_1999,burke_snakes_2007}, although for the range of parameters explored here, such type of structure has not been found.

The collapsed snaking structure is a consequence of the damped oscillatory interaction between the two DWs forming the LSs of type I (see Sec.~\ref{sec:3}). To understand this phenomenon let us take a look to the sketch shown in Fig.~\ref{DWs_LSs}(c). At the Maxwell point ($\nu=0$) a number stable and unstable LSs form at the stationary DWs separations $\Delta^n_s$. The stable (unstable) LSs in Fig.~\ref{DWs_LSs}(c) then correspond to a set of points on top of the stable (unstable) branches of solutions at $\rho_M$ in the collapsed snaking diagrams of Fig.~\ref{bif_diagram_typeI} [see for example the diagram shown in panel (f)]. As $\rho$ moves away from $\rho_M$, the branches of wider LSs start to disappear in a sequence of SN bifurcations, and only narrow LSs survive. At this point [see red vertical line in Fig.~\ref{bif_diagram_typeI}(f)] the scenario corresponds to the situation shown in Fig.~\ref{DWs_LSs}(c) for $\nu=\nu_1$, where four intersections of $f(D)$ with zero take place. Decreasing $\rho$ even further only two intersections occur [see Fig.~\ref{DWs_LSs}(c) for $\nu=\nu_2$] which correspond to the stable and unstable single peak  branches [see orange vertical line in Fig.~\ref{bif_diagram_typeI}(f)].

In this context, the SN bifurcations of the collapsed snaking diagram take place when the extrema of $f(D)$ become tangent to zero. Indeed, the tangency observed in Fig.~\ref{DWs_LSs}(c) corresponds to the occurrence of SN$_2^l$.
Eventually the last tangency corresponding to SN$_1^l$ occurs and the single peak LS is destroyed.

Decreasing $|\eta_2|$ to even lower values, the morphology of the collapsed snaking does not change much, despite the widening of the solution branches.As a result, the  the region of existence of the LSs increases [see Fig.~\ref{phase}]. This is the situation shown in Fig.~\ref{bif_diagram_typeI}(c)-(f) for $\eta_2=-0.05$. The LSs corresponding to this diagram are labeled with (xiii)-(xviii).


The widening of the LSs solution branches when decreasing $\eta_2$ is related with the modification of the oscillatory tails of the DWs involved in the formation of the LSs. It therefore depends directly on the spatial eigenvalues $\lambda$. Indeed, decreasing $|\eta_2|$ the oscillations in the tails become less damped, and its wavelength shortens. This can be appreciated when comparing the LSs plotted in panels (ii)-(vi) with those shown in (xiv)-(xviii).

The limit $\eta_2\rightarrow 0$ is particularly interesting. When $\eta_2=0$ the nonlocal nonlinear term becomes $\mathsf{A}^2\otimes \mathsf{J}=(1-i\Delta_2)\mathsf{A}^2$, and Eq.~(\ref{NL_sta}) reduces to  
\begin{equation}\label{GL}
\partial_t {\mathsf A}=-(1+i\Delta_1)\mathsf{A}-i\eta_1\partial_x^2\mathsf{A}-(1-i\Delta_2)|\mathsf{A}|^2\mathsf{A} +\rho\bar{\mathsf{A}},
\end{equation}
which is a particular version of the more general parametrically forced Ginzburg-Landau (PFGL) equation with 2:1 resonance, which has been studied in detail in \cite{burke_classification_2008}. We have confirmed, although not shown here, that the same type of solutions reported in this work are also present in model (\ref{GL}). Hence, the effect of $\eta_2$ mainly consists in modifying the region of existence of the type-I LSs, and eventually may imply their disappearance. 

While decreasing $\eta_2$ to zero, high-order dispersion terms may become relevant, and should normally be included in the study. The next term to be considered corresponds to the third-order dispersion effect. This term breaks the reflection symmetry $x\rightarrow -x$, inducing the drift of the LSs and the modification of the collapsed snaking as reported in \cite{parra-rivas_coexistence_2017}. Although these effects are very relevant regarding real physical systems, their study is beyond the scope of the present work, and will be examined elsewhere.

\section{Bifurcation structure of type-II localized states}\label{sec:6}
\begin{figure*}[t]
	\centering
	\includegraphics[scale=1.0]{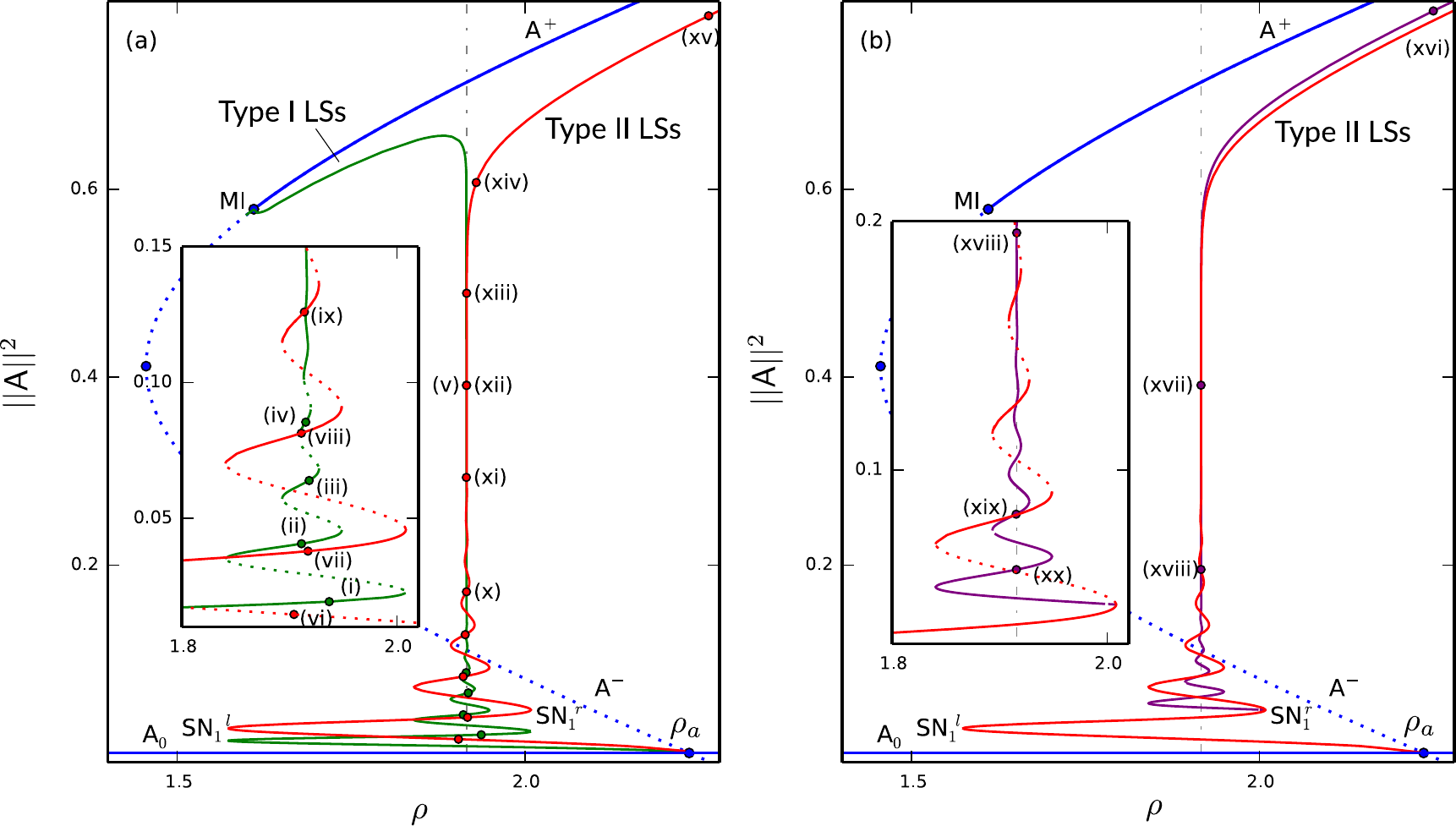}
		\includegraphics[scale=1.0]{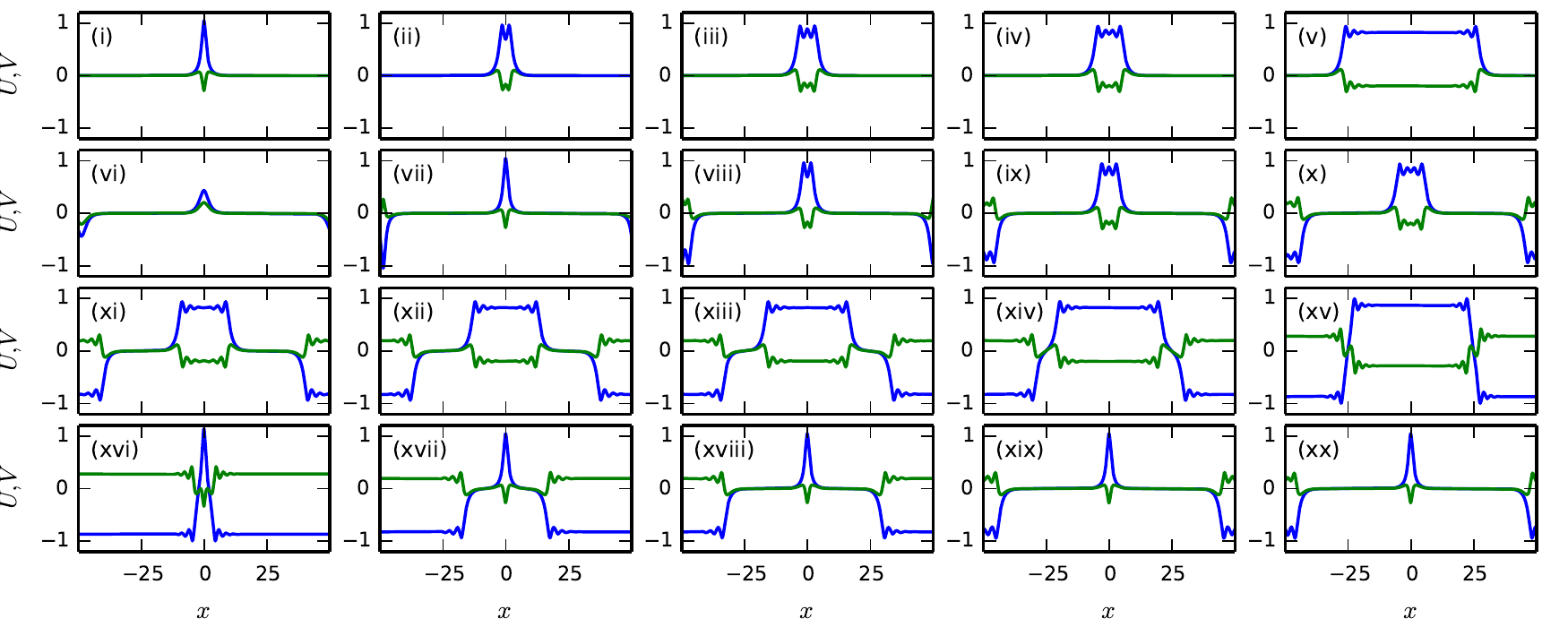}	
	\caption{(Color online) Bifurcation diagram for type-II LSs at $(\Delta_1,\eta_2)=(-2,-0.05)$. In panel (a) the collapsed snaking in green correspond to the type-I LSs [see profiles (i)-(v)] that has been added for comparison. The diagram in red correspond to the mixed structures shown in panels (vi)-(xiv) that eventually become a type-II LS as the one shown in panel (xv). The inset shows a close-up view of the bottom part of the bifurcation diagram including the stability of the branches, that alternates from unstable to stable between consecutive folds. I panel (b) we have removed the type-I bifurcation diagram, and added the purple diagram corresponding to transition shown in panels (xvi)-(xx).}
	\label{bif_diagram_typeII}
\end{figure*}
In this section we focus on the study of type-II LSs, and its bifurcation structure.  As discussed previously, these states are formed through the locking of DWs of different polarities connecting $-\mathsf{A}^+$ with $\mathsf{A}^+$. In contrast to the type-I states that exist in a reduced region around the Maxwell point, type-II LSs live in a broader area in parameter space
including regions II and III$_b$. When approaching $\rho_M$ in region II the type-II states become a hybrid state formed by two type-I LSs related by the symmetry $\mathsf{A}\rightarrow -\mathsf{A}$. In what follows we will show how this hybrid state also undergoes collapsed snaking. In this work we only consider stationary type-II LSs which are formed through the locking of DWs of Ising type \cite{coullet_breaking_1990}. For high values of $\rho$, the DWs may undergo non-equilibrium Ising-Bloch transitions \cite{coullet_breaking_1990}, resulting in the drifting of LSs, domain oscillations, and complex dynamics that were studied in detail in \cite{gomila_theory_2015}. 

To start we fix $(\Delta_1,\eta_2)=(-2,-0.05)$, as in the diagram shown in Fig.~\ref{bif_diagram_typeI}(c),(f), and we analyze the bifurcation structure associated with a hybrid state composed by two LSs of type-I which are related by the transformation $\mathsf{A}\rightarrow -\mathsf{A}$ and separated by half of the domain size $L/2$. The bifurcation structure corresponding to this type of states is shown red in Fig.~\ref{bif_diagram_typeII}(a). The bifurcation diagram in green is the same shown in Fig.~\ref{bif_diagram_typeI}(c)-(f) and is plotted here for comparison. The green dots correspond to the profiles labeled with (i)-(v).

Close to the pitchfork bifurcation $\rho_a$, solutions of the form $\mathsf{A}(x)-\mathsf{A}(x+L/2)$ exist, where $A(x)$ is the weakly nonlinear solution about $\rho_a$ (\ref{bump}). This mixed state corresponds to the profile (vi) plotted on the red curve shown in Fig.~\ref{bif_diagram_typeII}(a) [see close-up view]. When moving upwards along the curve, each component of this mixed state behaves as a single isolated state, undergoing collapsed snaking. At each fold on the right a new dip is nucleated from the center of each structure resulting in the widening of both states. This process can be seen in the profiles (ix)-(xi).
 
At this stage we can clearly identify the four DWs involved in the formation of the two LSs [see profile (xi)], which connect the CW solutions in the following sequence: $-\mathsf{A}^+\rightarrow \mathsf{A}_0\rightarrow \mathsf{A}^+\rightarrow \mathsf{A}_0\rightarrow -\mathsf{A}^+$. Increasing $||\mathsf{A}||^2$ further, the trivial state $\mathsf{A}_0$ decreases in width [see profiles (xii)-(xiv)] and eventually disappears. This occurs approximately at the moment that $\mathsf{A}_0$ becomes unstable. As a result the two DWs $-\mathsf{A}^+\rightarrow \mathsf{A}_0\rightarrow \mathsf{A}^+$ become a single DW connecting $-\mathsf{A}^+$ with $\mathsf{A}^+$, such that a pair of type-I LSs transforms into the single type-II state (see panel (xv)). This type-II state persists for higher values of $\rho$ and extends to region II. The linear stability of these structure is shown in the close-up view of Fig.~\ref{bif_diagram_typeII}(a).

We have verified that LSs of type-II with different initial widths undergo a similar type of bifurcation structure. To illustrate this behavior let us consider a single bump state, initially in region II, as the one plotted in panel (xvi). When modifying both $\rho$ and $||\mathsf{A}||^2$ this structure is described by the bifurcation diagram plotted in purple in Fig.~\ref{bif_diagram_typeII}(b), where we also plotted the bifurcation structure corresponding to the states (vi)-(xv) for comparison (red diagram).

When decreasing $\rho$, the LS (xvi) enters region III$_b$ where $\mathsf{A}_0$ is stable. Soon after that a plateau is created around $\mathsf{A}_0$ [see (xvii)] whose extension increases when approaching $\rho_M$ [see (xviii)]. At this stage one can clearly identify two DWs connecting $-\mathsf{A}^+$ with $\mathsf{A}_0$ and vice-versa, and the single bump type-II LSs becomes a pair of type-I LSs consisting in a bright bump sitting on $\mathsf{A}_0$ at the central position, and a dark wide LSs centered at distance $L/2$ from the former one. Proceeding down in the diagram the wider structure undergoes collapsed snaking losing one dip at each crossing of the SN$^r_i$ [see profiles (xviii)-(xx)], until it becomes just a dark single bump. This hybrid state finally collides with the red bifurcation diagram at SN$_1^r$ in symmetry-breaking pitchfork bifurcation. Indeed, at every SN$_i^r$ other pitchfork bifurcations occurs from where branches mixed states solutions emanates and undergo similar bifurcation structure, until becoming a type-II LS.

We have confirmed that for higher values of $|\eta_2|$ the bifurcation structure becomes much more complex, and therefore the numerical continuation of the LSs is more cumbersome. Despite of this complexity, the connection between type-I and II LSs persists and is qualitatively equivalent to the one shown in Fig.~\ref{bif_diagram_typeII}.

\section{Localized structures in the $(\Delta_1,\rho)-$parameter space}\label{sec:7}
\begin{figure}[t]
	\centering
	\includegraphics[scale=1]{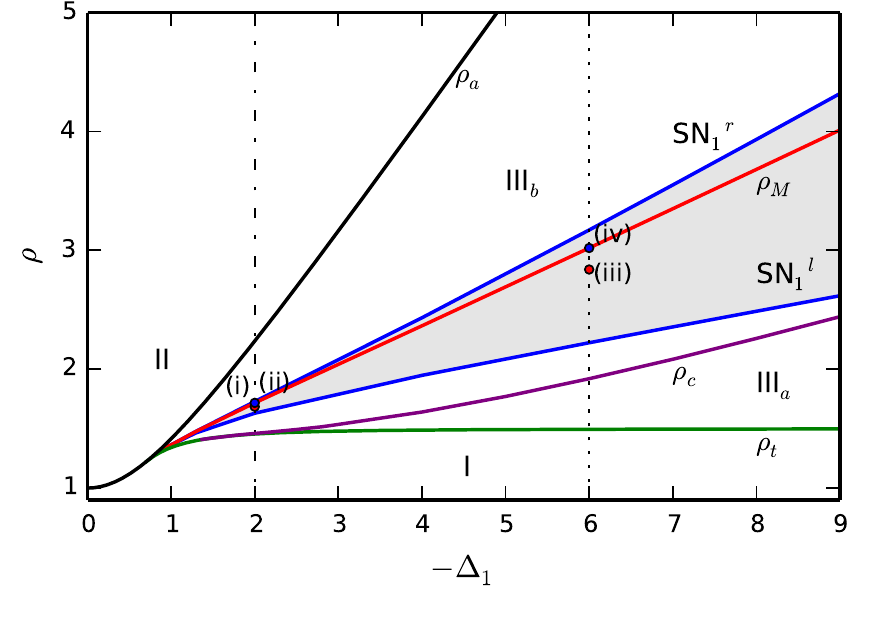}
		\includegraphics[scale=1]{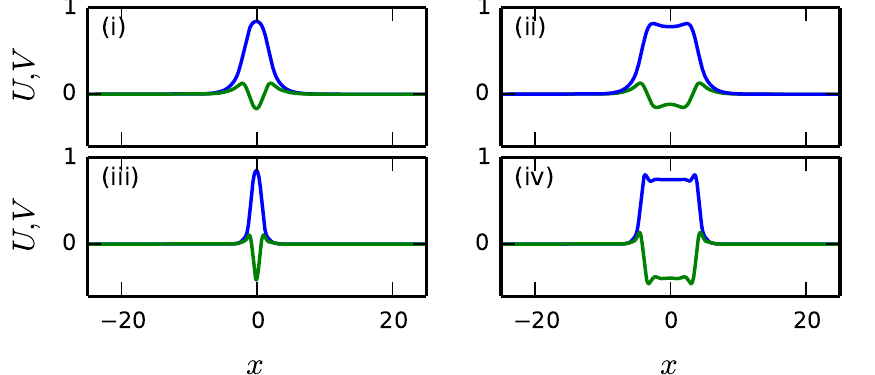}
	\caption{(Color online) Phase diagram in the $(\Delta_1,\rho)-$parameter space for $\eta_2=-0.4$. The gray area between SN$_1^l$ and SN$_1^r$ corresponds to the region where LSs of type-I exist. The pitchfork $\rho_a$ and saddle-node $\rho_t$ bifurcations of the CW solutions are plotted in black and green solid lines respectively. The MI is the purple line labeled by $\rho_c$, and the Maxwell point of the system $\rho_M$ is the red solid line. The inset shows a close-up of the phase diagram around the cusp bifurcation C. The labels (i)-(ii), and (iii)-(iv) correspond to the LSs shown in the panels below for $\Delta_1=-2$, and $\Delta_1=-6$ respectively.}
	\label{phase_theta}
\end{figure}

In previous sections we have fixed $\Delta_1=-2$ and studied how the different type of LSs and their bifurcation structure are modified when changing $\eta_2$. However, in experiments, $\eta_2$ is normally fixed when choosing the frequency of the input pump field, and $\Delta_1$ becomes one of the most relevant control parameter of the system. Because of that here we study the effect that the modification of $\Delta_1$ causes in the previously presented scenario when $\eta_2$ is fixed to $\eta_2=-0.4$. 

Figure~\ref{phase_theta} shows the phase diagram in the $(\Delta_1,\rho)-$ parameter space. Here, together with the pitchfork $\rho_a$ and saddle-node $\rho_t$ bifurcation lines, we have added the lines corresponding to SN$_1^l$ and SN$_1^r$ (blue curves), the Maxwell point $\rho_M$, and the MI $\rho_c$ corresponding to the chosen value $\eta_2=-0.4$. The gray area in-between SN$_1^l$ and SN$_1^r$ corresponds to the region where type-I DWs can lock and form LSs. 

When decreasing the absolute value of the $\Delta_1$, SN$_1^l$ and SN$_1^r$ approach one another and the gray region shrinks until it eventually disappears. SN$_1^l$ and SN$_1^r$ collide at the Maxwell point and disappear in a cusp bifurcation C. The Maxwell point then persists until $\Delta_1=1/\sqrt{2}$ where the SN$_t$ collides with $\mathsf{A}_0$ at $\rho_a$. 

In contrast, increasing $\Delta_1$ the region of existence widens, and as a result it is easier to find LSs. We find that LSs undergo the same type of collapsed snaking bifurcation diagram while modifying $\Delta_1$, what shows that these type of solutions and their bifurcation structure are robust. Panels (i)-(iv) show the LSs corresponding to two fixed values of $\Delta_1$: profiles (i)-(ii) correspond to $\Delta_1=-2$, and (iii)-(iv) to $\Delta_1=-6$ [see the dots on the vertical dashed lines in Fig.~\ref{phase_theta}]. 

In the limit of large $\Delta_1$, the mean field model (\ref{MF}) reduces to a single PFGL equation with pure Kerr nonlinearity that support analytical sech solutions of high amplitude \cite{longhi_localized_1997}. Those solutions would correspond in our work to the single bump type-I LS shown in panel (iii) for a large enough $\Delta_1$. However, no analytical solution has been found for the wider LS (iv).

Type-II LSs exist in region II and region III$_b$ for values of $\rho$ above $\rho_M$. However, for high values of $|\Delta_1|$ their bifurcation diagram can eventually become more complex.

\section{Discussion}\label{sec:8}
In this article we have presented a detailed and comprehensive analysis of the bifurcation structure and stability of LSs formed through locking of domain walls in $\chi^{(2)}-$dispersive cavities in the absence of walk-off. To do so we have considered a degenerate optical parametric oscillator in a doubly resonant configuration, and we have focused on the sub-critical regime.


To perform this analysis we have derived a PFGL type of equation with a nonlocal nonlinearity [see Eq.~(\ref{nl_GL})], which we have verified to reproduce the same results as the full mean-field model (\ref{MF}) (Sec.~\ref{sec:2}). In the PFGL context the pump field $B$ is dynamically slaved to $A$ [see Eq.~(\ref{B_slaved})], and therefore it is characterized by the latter.

\begin{figure}[t]
	\centering
	\includegraphics[scale=1]{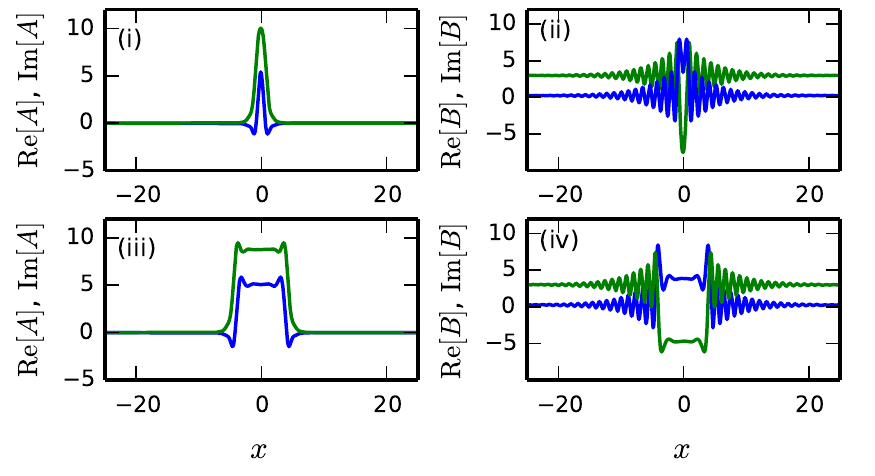}
	\caption{(Color online) Real and imaginary parts of of $A$ and $B$ for two different types of LSs of type-I. Panels (i)-(iii) show the signal field $A$, and panels (ii)-(iv) the correspondent pump field $B$. Here $(\Delta_1,\eta_2)=(-6,-4)$.}
	\label{TypeI_eta04}
\end{figure}

In regions II and III$_b$, the system is bistable and two types of DWs exist, forming connections between different CW solutions: i) $\mathsf{A}_0\rightarrow \mathsf{A}^+$, and  ii) $-\mathsf{A}^+\rightarrow \mathsf{A}^+$. We referred to these DWs as type-I and type-II. In the presence of oscillatory tails, two DWs with different polarities can lock forming LSs of different widths. We refer to these LSs as type-I and II, depending on the type of DW involved in their formation. We have shown that LSs of type-I undergo collapsed snaking \cite{knobloch_homoclinic_2005,ma_defect-mediated_2010,parra-rivas_dark_2016}. Here "collapsed" refers to the fact that the region of existence of LSs shrinks exponentially as the width of the LS increases. Wider structures can only be found around the Maxwell point, and the observation of LSs with a single bump is favored. Two examples of such type-I LSs are plotted in Fig.~\ref{TypeI_eta04} for $(\Delta_1,\eta_2)=(-6,-4)$ using the variable $A$ and $B$: Panel (i) shows the real and imaginary part of $A$ for a single bump LS, and in panel (ii) the slaved field $B$ is plotted using the relation (\ref{B_slaved}). Panels (iii) and (iv) represent the pump and signal fields corresponding to a wide structure. 

The collapsed snaking emerges from the pitchfork bifurcation on $\mathsf{A}_0$, and it connects back to the non-trivial CW state $\mathsf{A}^+$, either at the saddle-node SN$_t$, or at the MI, depending on the control parameters of the system. Applying multi-scale perturbation methods, we have been able to calculate 
an analytical sech pulse-like solution close to the pitchfork bifurcation at $\rho_a$.

We have studied how the LSs and their associated bifurcation structure are modified when the group velocity dispersion $\eta_2$ changes. For doing so we fixed $\Delta_1=-2$ and calculated the phase diagram in the $(\eta_2,\rho)-$parameter space shown in Fig.~\ref{phase}.
The phase diagram shows that when increasing $|\eta_2|$ the type-I LSs disappear, while type-II LSs persist in region II and III$_b$ well above the Maxwell point. In contrast, decreasing $|\eta_2|$, the region of existence of type-I LSs increases, and many more type-I LSs can be found. 

When $\eta_2\rightarrow 0$, the nonlocal nonlinear model (\ref{nl_GL}) reduces to Eq.~(\ref{GL}), which is a simpler case of the more general PFGL equation \cite{burke_classification_2008}. We have confirmed that the LSs presented in this work persist in such limit, and undergo a similar type of bifurcation structure. A complete understanding of this PFGL system [Eq.~(\ref{GL})] is of great interest, and a detailed study of this model will be presented elsewhere. 
From a physical perspective the previous limit must be considered carefully since when $|\eta_2|$ becomes very small, high-order dispersion effects may play an essential role and should be taken into account.

In addition to the type-I LSs, a large variety of type-II LSs also formed through locking of DWs connecting the equivalent states $-\mathsf{A}^+$ with $\mathsf{A}^+$. These states exist for a wider range of parameters in region II and III$_b$, and may undergo non-equilibrium Ising-Bloch transitions \cite{coullet_breaking_1990}, resulting in complex dynamics \cite{gomila_theory_2015}. We have shown that in region III$_b$, every type-II LS becomes a hybrid state composed by two type-I LSs related by the symmetry $\mathsf{A}\rightarrow -\mathsf{A}$ and are separated by $L/2$. Moreover, each of these states independently undergoes collapsed snaking around the Maxwell point, which is also the bifurcation structure characterizing its components.

Finally, in Sec.~\ref{sec:7}, we have shown that type-I and II LSs persist for different values of $\Delta_1$, and that they are described by the same kind of bifurcation structure.

\section{Conclusion}\label{sec:9}
The analysis presented in this paper provides a detailed study of the bifurcation structure and stability of the LSs arising in doubly resonant optical parametric oscillators in the absence of temporal walk-off. A potential physical realizable configuration for which the walk-off vanishes is described in \cite{hansson_quadratic_2018}.

 The type of states studied here arise through the locking of DWs formed between two continuous wave states that coexist in the same parameter range, i.e. in the presence of bistability. The oscillatory damped nature of the DW interaction determines a particular bifurcation structure known as {\it collapsed snaking}, which is generic and appears in a large number of systems in different contexts \cite{knobloch_homoclinic_2005,ma_defect-mediated_2010,burke_classification_2008,parra-rivas_dark_2016}. In contrast to the type-II LSs, which have been analyzed in detail in quadratic cavities \cite{oppo_domain_1999,
oppo_characterization_2001}, as far as we known, the type-I LSs presented here have not been reported elsewhere.

To perform this analysis we have derived a nonlinear nonlocal model (\ref{nl_GL}) similar to those derived for quadratic nonlinear cavities \cite{leo_walk-off-induced_2016,leo_frequency-comb_2016,mosca_modulation_2018}. The results found here can be extended to singly resonant cavities, where the model is formally equivalent to (\ref{nl_GL}), albeit with a different nonlocal response \cite{mosca_modulation_2018}.

A natural extension of this work must include the effect of the temporal walk-off, which breaks the $x\rightarrow-x$ symmetry inducing asymmetry and drift. We expect that for weak walk-off the collapsed snaking is modified in the same fashion as in the context of Kerr cavities in the presence of third-order dispersion \cite{parra-rivas_coexistence_2017}.

Quadratic dispersive cavities have gained a lot interest in the past few years as an alternative to Kerr cavities for the generation of optical frequency combs \cite{leo_walk-off-induced_2016,leo_frequency-comb_2016, mosca_frequency_2017,mosca_modulation_2018,hansson_quadratic_2018}. Therefore, these results present a series of wave-forms whose frequency spectrum could be of interest for applications.

\acknowledgments 
We acknowledge the support from internal Funds from KU Leuven and the FNRS (PPR), and funding from the European Research Council (ERC) under the European Union’s Horizon 2020 research and innovation programme [Grant agreement No. 757800], (FL).

\section*{Appendix A: Derivation of the parametrically forced Ginzburg-Landau equation with nonlinear nonlocal coupling}\label{Appen:A}
In this Appendix we derive the Eq.~(\ref{nl_GL}) from the mean-field model (\ref{MF}). To do so we apply the same procedure than in Refs.~\cite{nikolov_quadratic_2003,leo_frequency-comb_2016}. 
This approach assumes the adiabatic elimination of the pump field $B$ in Eq.~(\ref{MF2}), i.e. $B$ varies slowly with $t$, at least at time scale slower than the $A$ field.
Hence, one can assume that $\partial_tB\approx0$ and thus Eq.~(\ref{MF2}) reduces to 
\begin{equation}\label{adia}
-\left(\alpha+i\Delta_2+d\partial_x+i\eta_2\partial_x^2\right)B+iA^2+S=0.
\end{equation}
Defining the direct and inverse Fourier transforms
\begin{equation}
\mathcal{F}[f(x)](k)=\int_{-\infty}^{\infty}e^{ikx}f(x)dx=\tilde{f}(k),
\end{equation}
and
\begin{equation}
\mathcal{F}^{-1}[\tilde{f}(k)](x)=\frac{1}{2\pi}\int_{-\infty}^{\infty}e^{-ikx}\tilde{f}(k)dk= f(x),
\end{equation}
one gets from Eq.~(\ref{adia}) 
\begin{equation}\label{Fourier_B}
\mathcal{F}[B]=i\mathcal{F}[J]\mathcal{F}[A^2]+\mathcal{F}[J]\mathcal{F}[S]
\end{equation}
with
\begin{equation}
\mathcal{F}[J(k)]=\frac{1}{\alpha+i(\Delta_2+k d-\eta_2k^2)}.
\end{equation}
Applying the inverse Fourier transform, Eq.~(\ref{Fourier_B}) then becomes
\begin{equation}\label{B1}
B(x)=i\mathcal{F}^{-1}\left(\mathcal{F}(J)\cdot\mathcal{F}(A^2)\right)+\mathcal{F}^{-1}\left(\mathcal{F}(J)\cdot\mathcal{F}(S)\right).
\end{equation}
Due to the convolution theorem, the first term on the right-hand side (rhs) of Eq.~(\ref{B1}) becomes
\begin{equation}
\mathcal{F}^{-1}\left[\mathcal{F}(J)\cdot\mathcal{F}(A^2)\right]=\int_{-\infty}^{\infty}J(x')A^2(x-x')dx'=J\otimes A^2,
\end{equation}
where $J(x)$ is the kernel defining a long-range nonlocal coupling in $x$, and $\otimes$ stands for the convolution operation. 

With the definition of Dirac distribution 
\begin{equation}
\delta(k)=\frac{1}{2\pi}\int_{-\infty}^{\infty}e^{ikx}dx,
\end{equation}
and taking $\tilde\Delta_2=\Delta_2/\alpha$, the second term on (rhs) of Eq.~(\ref{B1}) yields  
\begin{multline}
\mathcal{F}^{-1}\left[\mathcal{F}(J)\cdot\mathcal{F}(S)\right]=2\pi S\mathcal{F}^{-1}\left[\mathcal{F}(J)\cdot \delta(k)\right]=\\S\mathcal{F}[J(0)]=
\frac{S}{\alpha\sqrt{1+\tilde\Delta_2^2}}e^{i{\rm atan}(-\tilde\Delta_2)}
\end{multline}
Thus the pump field finally reads, 
\begin{equation}\label{B_field}
B(x)=iJ\otimes A^2+\rho e^{i{\rm atan}(-\tilde\Delta_2)},
\end{equation}
where we have defined 
\begin{equation}
\rho=\frac{S}{\alpha\sqrt{1+\tilde\Delta_2^2}}.
\end{equation}
Inserting (\ref{B_field}) into Eq.~(\ref{MF1}), the later becomes in PFGL type of equation with a nonlinear nonlocal long range interaction term:
\begin{equation}\label{NLGL_eq}
\partial_t A=-(1+i\Delta_1)A-i\eta_1\partial_x^2 A-\bar{A}(J\otimes A^2)+\rho\bar{A}e^{2i\psi},
\end{equation}
with
\begin{equation}
\psi=\pi/4+{\rm atan}(-\tilde\Delta_2)/2.
\end{equation}
Rescaling the A field as 
\begin{equation}\label{normalized_A}
A=\mathsf{A}e^{i\psi}\sqrt{\alpha(1+\tilde\Delta_2^2)}
\end{equation}
the Eq.~(\ref{NLGL_eq}) then becomes
\begin{equation}\label{NLGL_normalized}
\partial_t {\mathsf A}=-(1+i\Delta_1)\mathsf{A}-i\eta_1\partial_{x}^2\mathsf{A}-\bar{\mathsf{A}}(\mathsf{A}^2\otimes\mathsf{J}) +\rho\bar{\mathsf{A}}.
\end{equation}				
With this normalization the long-range interaction kernel becomes 
\begin{equation}
\mathsf{J}(x)=\frac{1+\tilde\Delta_2^2}{2\pi}\int_{-\infty}^{\infty}\frac{e^{-ik x}dx}{1+i(\tilde\Delta_2+\gamma k-\tilde\eta_2 k^2)},
\end{equation}
with $\gamma=d/\alpha$, and $\tilde\eta_2=\eta_2/\alpha$.										
The real and imaginary parts of this kernel are
\begin{equation}\label{kernel_R}
\mathsf{J}_R(x)=\frac{1+\tilde\Delta_2^2}{2\pi}\int_{-\infty}^{\infty}\frac{e^{-ik x}dk}{1+(\tilde\Delta_2+\gamma k-\tilde\eta_2 k^2)^2},
\end{equation}

\begin{equation}\label{kernel_I}
\mathsf{J}_I(x)=-\frac{1+\tilde\Delta_2^2}{2\pi}\int_{-\infty}^{\infty}\frac{(\tilde\Delta_2+\gamma k-\tilde\eta_2 k^2)e^{-ik x}dx}{1+(\tilde\Delta_2+\gamma k-\tilde\eta_2 k^2)^2}.
\end{equation}
With this normalization the B field becomes
\begin{equation}
B=(-\mathsf{A}^2\otimes\mathsf{J}+\rho)e^{i{\rm atan}(-\tilde\Delta_2)}
\end{equation}
In this work we consider $\gamma=0$, and therefore (\ref{kernel_R}) and (\ref{kernel_I}) are symmetric under the transformation $x\rightarrow-x$. 
The square root factor in Eq.~(\ref{normalized_A}) has been introduced for convenience in order to obtain the standard form of the PFGL Eq.~(\ref{GL}) in the limit $\gamma,\tilde\eta_2\rightarrow 0$.

\section*{Appendix B: Weakly nonlinear analysis around the Pitchfork bifurcation}\label{Appen:B}
In this Appendix we show how to obtain the stationary amplitude equation (\ref{normal_form}) around $\rho_a$ starting from the equation at order $\epsilon^3$ in the perturbation expansion, namely:
\begin{equation}
\mathcal{L}_0\left[\begin{array}{c}
u_3 \\ v_3
\end{array}
\right]=-(\mathcal{L}_2+\mathcal{N}_2)\left[\begin{array}{c}
u_1 \\ v_1
\end{array}
\right].
\end{equation}
To solve this equation we have first to deal with the nonlinear nonlocal operator 
\begin{equation}\label{nonl_2order}
\mathcal{N}=\epsilon^2\mathcal{N}_2=-\left[\begin{array}{cc}
\mathcal{N}_2^a & \mathcal{N}_2^b\\
\mathcal{N}_2^b & -\mathcal{N}_2^a
\end{array}\right],
\end{equation}
with 
\begin{subequations}
	\begin{equation}
	\mathcal{N}_2^a=u_1^2\otimes J_R-v_1^2\otimes J_R-2u_1v_1\otimes J_I,  
	\end{equation}
	\begin{equation}
	\mathcal{N}_2^b= u_1^2\otimes J_I-v_1^2\otimes J_I+2u_1v_1\otimes J_R,
	\end{equation}
\end{subequations}	
where the solution of the problem at $\mathcal{O}(\epsilon)$ reads
	\begin{equation}
	\left[\begin{array}{c}
	u_1 \\ v_1
	\end{array}\right]=\left[\begin{array}{c}
	\xi \\ 1
	\end{array}\right]a(x_1),
	\end{equation}
	with $\xi=\Delta_1/(1-\rho_a)$ and $a(x_1)$ a real function.
	
At this point we have to evaluate the convolution terms, and for doing so we follow the procedure described in Ref.~\cite{morgan_swifthohenberg_2014,kuehn_validity_2018}. In order to perform this calculation we consider that all the terms posed on the long length scale $x_1$ are assumed to be almost constant over the region where the kernel $J$ is large, what is equivalent to consider for a very narrow kernel \citep{krolikowski_modulational_2001}. This makes sense when one assumes that the amplitude $a$ of the envelope is smooth, and the kernel decays much more rapidly than the envelope.
	
With these considerations we obtain:
	\begin{subequations}
		\begin{multline*}
		u_1^2\otimes J_R=\int_{-\infty}^{\infty} u_1^2(x')J_R(x-x')dx'=\\ \xi^2\int_{-\infty}^{\infty} a(x'_1)^2J_R(x-x')dx'\approx\xi^2 a(x_1)^2\int_{-\infty}^{\infty}J_R(x-x')dx'=\\
		\xi^2 a(x_1)^2\mathcal{F}^{-1}\left(2\pi\delta(k)\mathcal{F}[J_R](k)\right)=\\\xi^2 a(x_1)^2\mathcal{F}[J_R](0)=\xi^2 a(x_1)^2,
		\end{multline*}
		\begin{multline*}
		u_1v_1\otimes J_I=\int_{-\infty}^{\infty} u_1(x')v_1(x')J_I(x-x')dx'\\\approx\xi a(x_1)^2\int_{\infty}^{\infty}J_I(x-x')dx'=\\
		\xi a(x_1)^2\mathcal{F}[J_I](0)=-\xi\Delta_2 a(x_1)^2,
		\end{multline*}
\end{subequations}
\begin{subequations}
		and with the same approach
		\begin{equation*}
		v_1^2\otimes J_R\approx a(x_1)^2\mathcal{F}[J_R](0)=a(x_1)^2,
		\end{equation*}
		
		\begin{equation*}
		u_1^2\otimes J_I\approx -\Delta_2\xi^2 a(x_1)^2,
		\end{equation*}
		\begin{equation*}
		v_1^2\otimes J_I\approx-\Delta_2 a(x_1)^2,
		\end{equation*}
		\begin{equation*}
		u_1v_1\otimes J_R\approx\xi a(x_1)^2,
		\end{equation*}
	\end{subequations}
	Thus, the the components of the nonlinear operator (\ref{nonl_2order}) become 
	\begin{subequations}
	\begin{equation}
	\mathcal{N}_2^a=(\xi^2+2\xi\Delta_2-1)a(x_1)^2
	\end{equation}
		\begin{equation}
		\mathcal{N}_2^b=(-\xi^2\Delta_2+2\xi+\Delta_2)a(x_1)^2
		\end{equation}
	\end{subequations}
	The amplitude equation about $\rho_a$ is then obtained from the solvability condition 
\begin{equation}
w^T\cdot\mathcal{L}_2\left[\begin{array}{c}
u_1 \\ v_1
\end{array}
\right]+w^T\cdot\mathcal{N}_2\left[\begin{array}{c}
u_1\\v_1
\end{array}\right]=\left[\begin{array}{c}
0\\0\end{array}\right],\end{equation}
where $w^T=[-\xi, 1]$, such that $\mathcal{L}^{\dagger}_0w=0$.	
 
The evaluation of the first term yields   
\begin{equation}
w^T\cdot\mathcal{L}_2\left[\begin{array}{c} u_1\\v_1
\end{array}\right]=-\delta(\xi^2+1) a-2\xi\eta_1\partial^2_{x_1} a,
\end{equation}
	while the second one gives	
				\begin{equation}
				w^T\cdot\mathcal{N}_2\left[\begin{array}{c} u_1\\v_1
				\end{array}\right]=(\xi^2+1)(\xi^2+2\Delta_2\xi-1)a(x_1)^3.
				\end{equation}
After arranging these terms and simplifying them one gets the stationary amplitude equation (\ref{normal_form}) about $\rho_a$.
\bibliographystyle{ieeetr}
\bibliography{DOPOs_LSs_v2}

\begin{thebibliography}{10}

\bibitem{coullet_nature_1987}
P.~Coullet, C.~Elphick, and D.~Repaux, ``Nature of spatial chaos,'' {\em
  Physical Review Letters}, vol.~58, pp.~431--434, Feb. 1987.

\bibitem{coullet_localized_2002}
P.~Coullet, ``Localized patterns and fronts in nonequilibrium systems,'' {\em
  International Journal of Bifurcation and Chaos}, vol.~12, pp.~2445--2457,
  Nov. 2002.

\bibitem{tlidi_localized_1994}
M.~Tlidi, P.~Mandel, and R.~Lefever, ``Localized structures and localized
  patterns in optical bistability,'' {\em Physical Review Letters}, vol.~73,
  pp.~640--643, Aug. 1994.

\bibitem{fernandez-oto_c._strong_2014}
{Fernandez-Oto C.}, {Tlidi M.}, {Escaff D.}, and {Clerc M. G.}, ``Strong
  interaction between plants induces circular barren patches: fairy circles,''
  {\em Philosophical Transactions of the Royal Society A: Mathematical,
  Physical and Engineering Sciences}, vol.~372, p.~20140009, Oct. 2014.

\bibitem{ruiz-reynes_fairy_2017}
D.~Ruiz-Reynés, D.~Gomila, T.~Sintes, E.~Hernández-García, N.~Marbà, and
  C.~M. Duarte, ``Fairy circle landscapes under the sea,'' {\em Science
  Advances}, vol.~3, p.~e1603262, Aug. 2017.

\bibitem{willebrand_experimental_1993}
H.~Willebrand, M.~Or-Guil, M.~Schilke, H.~G. Purwins, and Y.~A. Astrov,
  ``Experimental and numerical observation of quasiparticle like structures in
  a distributed dissipative system,'' {\em Physics Letters A}, vol.~177,
  pp.~220--224, June 1993.

\bibitem{umbanhowar_localized_1996}
P.~B. Umbanhowar, F.~Melo, and H.~L. Swinney, ``Localized excitations in a
  vertically vibrated granular layer,'' {\em Nature}, vol.~382, p.~793, Aug.
  1996.

\bibitem{taranenko_patterns_2000}
V.~B. Taranenko, I.~Ganne, R.~J. Kuszelewicz, and C.~O. Weiss, ``Patterns and
  localized structures in bistable semiconductor resonators,'' {\em Physical
  Review A}, vol.~61, p.~063818, May 2000.

\bibitem{ramazza_localized_2000}
P.~L. Ramazza, S.~Ducci, S.~Boccaletti, and F.~T. Arecchi, ``Localized versus
  delocalized patterns in a nonlinear optical interferometer,'' {\em Journal of
  Optics B: Quantum and Semiclassical Optics}, vol.~2, pp.~399--405, June 2000.

\bibitem{barland_cavity_2002}
S.~Barland, J.~R. Tredicce, M.~Brambilla, L.~A. Lugiato, S.~Balle, M.~Giudici,
  T.~Maggipinto, L.~Spinelli, G.~Tissoni, T.~Knödl, M.~Miller, and R.~Jäger,
  ``Cavity solitons as pixels in semiconductor microcavities,'' {\em Nature},
  vol.~419, p.~699, Oct. 2002.

\bibitem{leo_temporal_2010}
F.~Leo, S.~Coen, P.~Kockaert, S.-P. Gorza, P.~Emplit, and M.~Haelterman,
  ``Temporal cavity solitons in one-dimensional {Kerr} media as bits in an
  all-optical buffer,'' {\em Nature Photonics}, vol.~4, pp.~471--476, July
  2010.

\bibitem{nicolis_self-organization_1977}
G.~Nicolis and I.~Prigogine, {\em Self-organization in nonequilibrium systems:
  from dissipative structures to order through fluctuations}.
\newblock New York, N.Y.: Wiley, 1977.
\newblock OCLC: 797228045.

\bibitem{akhmediev_dissipative_2005}
N.~Akhmediev and A.~Ankiewicz, eds., {\em Dissipative {Solitons}}.
\newblock Lecture {Notes} in {Physics}, Berlin Heidelberg: Springer-Verlag,
  2005.

\bibitem{murray_mathematical_2002}
J.~D. Murray, {\em Mathematical {Biology}: {I}. {An} {Introduction}}.
\newblock Interdisciplinary {Applied} {Mathematics}, {Mathematical} {Biology},
  New York: Springer-Verlag, 3~ed., 2002.

\bibitem{clerc_patterns_2005}
M.~G. Clerc, D.~Escaff, and V.~M. Kenkre, ``Patterns and localized structures
  in population dynamics,'' {\em Physical Review E}, vol.~72, p.~056217, Nov.
  2005.

\bibitem{scroggie_pattern_1994}
A.~J. Scroggie, W.~J. Firth, G.~S. McDonald, M.~Tlidi, R.~Lefever, and L.~A.
  Lugiato, ``Pattern formation in a passive {Kerr} cavity,'' {\em Chaos,
  Solitons \& Fractals}, vol.~4, pp.~1323--1354, Aug. 1994.

\bibitem{firth_two-dimensional_1996}
W.~J. Firth and A.~Lord, ``Two-dimensional solitons in a {Kerr} cavity,'' {\em
  Journal of Modern Optics}, vol.~43, pp.~1071--1077, May 1996.

\bibitem{etrich_solitary_1997}
C.~Etrich, U.~Peschel, and F.~Lederer, ``Solitary {Waves} in {Quadratically}
  {Nonlinear} {Resonators},'' {\em Physical Review Letters}, vol.~79,
  pp.~2454--2457, Sept. 1997.

\bibitem{staliunas_localized_1997}
K.~Staliunas and V.~J. Sánchez-Morcillo, ``Localized structures in degenerate
  optical parametric oscillators,'' {\em Optics Communications}, vol.~139,
  pp.~306--312, July 1997.

\bibitem{staliunas_spatial-localized_1998}
K.~Staliunas and V.~J. Sánchez-Morcillo, ``Spatial-localized structures in
  degenerate optical parametric oscillators,'' {\em Physical Review A},
  vol.~57, pp.~1454--1457, Feb. 1998.

\bibitem{longhi_localized_1997}
S.~Longhi, ``Localized structures in optical parametric oscillation,'' {\em
  Physica Scripta}, vol.~56, pp.~611--618, Dec. 1997.

\bibitem{oppo_domain_1999}
G.-L. Oppo, A.~J. Scroggie, and W.~J. Firth, ``From domain walls to localized
  structures in degenerate optical parametric oscillators,'' {\em Journal of
  Optics B: Quantum and Semiclassical Optics}, vol.~1, pp.~133--138, Jan. 1999.

\bibitem{oppo_characterization_2001}
G.-L. Oppo, A.~J. Scroggie, and W.~J. Firth, ``Characterization, dynamics and
  stabilization of diffractive domain walls and dark ring cavity solitons in
  parametric oscillators,'' {\em Physical Review E}, vol.~63, May 2001.

\bibitem{staliunas_transverse_2003}
K.~Staliunas and V.~J. Sánchez-Morcillo, {\em Transverse {Patterns} in
  {Nonlinear} {Optical} {Resonators}}.
\newblock Springer {Tracts} in {Modern} {Physics}, Berlin Heidelberg:
  Springer-Verlag, 2003.

\bibitem{chembo_spatiotemporal_2013}
Y.~K. Chembo and C.~R. Menyuk, ``Spatiotemporal {Lugiato}-{Lefever} formalism
  for {Kerr}-comb generation in whispering-gallery-mode resonators,'' {\em
  Physical Review A}, vol.~87, p.~053852, May 2013.

\bibitem{leo_dynamics_2013}
F.~Leo, L.~Gelens, P.~Emplit, M.~Haelterman, and S.~Coen, ``Dynamics of
  one-dimensional {Kerr} cavity solitons,'' {\em Optics Express}, vol.~21,
  pp.~9180--9191, Apr. 2013.

\bibitem{herr_temporal_2014}
T.~Herr, V.~Brasch, J.~D. Jost, C.~Y. Wang, N.~M. Kondratiev, M.~L. Gorodetsky,
  and T.~J. Kippenberg, ``Temporal solitons in optical microresonators,'' {\em
  Nature Photonics}, vol.~8, pp.~145--152, Feb. 2014.

\bibitem{delhaye_optical_2007}
P.~Del’Haye, A.~Schliesser, O.~Arcizet, T.~Wilken, R.~Holzwarth, and T.~J.
  Kippenberg, ``Optical frequency comb generation from a monolithic
  microresonator,'' {\em Nature}, vol.~450, pp.~1214--1217, Dec. 2007.

\bibitem{kippenberg_microresonator-based_2011}
T.~J. Kippenberg, R.~Holzwarth, and S.~A. Diddams, ``Microresonator-{Based}
  {Optical} {Frequency} {Combs},'' {\em Science}, vol.~332, pp.~555--559, Apr.
  2011.

\bibitem{pasquazi_micro-combs:_2018}
A.~Pasquazi, M.~Peccianti, L.~Razzari, D.~J. Moss, S.~Coen, M.~Erkintalo, Y.~K.
  Chembo, T.~Hansson, S.~Wabnitz, P.~Del’Haye, X.~Xue, A.~M. Weiner, and
  R.~Morandotti, ``Micro-combs: {A} novel generation of optical sources,'' {\em
  Physics Reports}, vol.~729, pp.~1--81, Jan. 2018.

\bibitem{leo_walk-off-induced_2016}
F.~Leo, T.~Hansson, I.~Ricciardi, M.~De~Rosa, S.~Coen, S.~Wabnitz, and
  M.~Erkintalo, ``Walk-{Off}-{Induced} {Modulation} {Instability}, {Temporal}
  {Pattern} {Formation}, and {Frequency} {Comb} {Generation} in
  {Cavity}-{Enhanced} {Second}-{Harmonic} {Generation},'' {\em Physical Review
  Letters}, vol.~116, p.~033901, Jan. 2016.

\bibitem{leo_frequency-comb_2016}
F.~Leo, T.~Hansson, I.~Ricciardi, M.~De~Rosa, S.~Coen, S.~Wabnitz, and
  M.~Erkintalo, ``Frequency-comb formation in doubly resonant second-harmonic
  generation,'' {\em Physical Review A}, vol.~93, p.~043831, Apr. 2016.

\bibitem{mosca_frequency_2017}
S.~Mosca, M.~Parisi, I.~Ricciardi, F.~Leo, T.~Hansson, M.~Erkintalo,
  P.~Maddaloni, P.~D. Natale, S.~Wabnitz, S.~Wabnitz, and M.~D. Rosa,
  ``Frequency comb generation in continuously pumped optical parametric
  oscillator,'' in {\em Frontiers in {Optics} 2017 (2017), paper {FTh}2B.4},
  p.~FTh2B.4, Optical Society of America, Sept. 2017.

\bibitem{mosca_modulation_2018}
S.~Mosca, M.~Parisi, I.~Ricciardi, F.~Leo, T.~Hansson, M.~Erkintalo,
  P.~Maddaloni, P.~De~Natale, S.~Wabnitz, and M.~De~Rosa, ``Modulation
  {Instability} {Induced} {Frequency} {Comb} {Generation} in a {Continuously}
  {Pumped} {Optical} {Parametric} {Oscillator},'' {\em Physical Review
  Letters}, vol.~121, p.~093903, Aug. 2018.

\bibitem{hansson_quadratic_2018}
T.~Hansson, P.~Parra-Rivas, M.~Bernard, F.~Leo, L.~Gelens, and S.~Wabnitz,
  ``Quadratic soliton combs in doubly resonant second-harmonic generation,''
  {\em Optics Letters}, vol.~43, pp.~6033--6036, Dec. 2018.

\bibitem{trillo_stable_1997}
S.~Trillo, M.~Haelterman, and A.~Sheppard, ``Stable topological spatial
  solitons in optical parametric oscillators,'' {\em Optics Letters}, vol.~22,
  pp.~970--972, July 1997.

\bibitem{parra-rivas_frequency_2019}
P.~Parra-Rivas, L.~Gelens, T.~Hansson, S.~Wabnitz, and F.~Leo, ``Frequency comb
  generation through the locking of domain walls in doubly resonant dispersive
  optical parametric oscillators,'' {\em Optics Letters}, vol.~44,
  pp.~2004--2007, Apr. 2019.

\bibitem{haelterman_dissipative_1992}
M.~Haelterman, S.~Trillo, and S.~Wabnitz, ``Dissipative modulation instability
  in a nonlinear dispersive ring cavity,'' {\em Optics Communications},
  vol.~91, pp.~401--407, Aug. 1992.

\bibitem{oppo_formation_1994}
G.-L. Oppo, M.~Brambilla, and L.~A. Lugiato, ``Formation and evolution of roll
  patterns in optical parametric oscillators,'' {\em Physical Review A},
  vol.~49, pp.~2028--2032, Mar. 1994.

\bibitem{zambrini_convection-induced_2005}
R.~Zambrini, M.~San~Miguel, C.~Durniak, and M.~Taki, ``Convection-induced
  nonlinear symmetry breaking in wave mixing,'' {\em Physical Review E},
  vol.~72, p.~025603, Aug. 2005.

\bibitem{nikolov_quadratic_2003}
N.~I. Nikolov, D.~Neshev, O.~Bang, and W.~Z. Królikowski, ``Quadratic solitons
  as nonlocal solitons,'' {\em Physical Review E}, vol.~68, Sept. 2003.

\bibitem{burke_classification_2008}
J.~Burke, A.~Yochelis, and E.~Knobloch, ``Classification of {Spatially}
  {Localized} {Oscillations} in {Periodically} {Forced} {Dissipative}
  {Systems},'' {\em SIAM Journal on Applied Dynamical Systems}, vol.~7,
  pp.~651--711, Jan. 2008.

\bibitem{lin_raman_2006}
Q.~Lin and G.~P. Agrawal, ``Raman response function for silica fibers,'' {\em
  Optics Letters}, vol.~31, pp.~3086--3088, Nov. 2006.

\bibitem{chembo_spatiotemporal_2015}
Y.~K. Chembo, I.~S. Grudinin, and N.~Yu, ``Spatiotemporal dynamics of
  {Kerr}-{Raman} optical frequency combs,'' {\em Physical Review A}, vol.~92,
  p.~043818, Oct. 2015.

\bibitem{krolikowski_modulational_2001}
W.~Z. Krolikowski, O.~Bang, W.~Krolikowski, and J.~Wyller, ``Modulational
  instability in nonlocal nonlinear {Kerr} media.,'' {\em Physical review. E,
  Statistical, nonlinear, and soft matter physics}, vol.~64, no.~1, p.~016612,
  2001.

\bibitem{suter_stabilization_1993}
D.~Suter and T.~Blasberg, ``Stabilization of transverse solitary waves by a
  nonlocal response of the nonlinear medium,'' {\em Physical Review A},
  vol.~48, pp.~4583--4587, Dec. 1993.

\bibitem{firth_proposed_2007}
W.~J. Firth, L.~Columbo, and A.~J. Scroggie, ``Proposed {Resolution} of
  {Theory}-{Experiment} {Discrepancy} in {Homoclinic} {Snaking},'' {\em
  Physical Review Letters}, vol.~99, p.~104503, Sept. 2007.

\bibitem{gordon_longtransient_1965}
J.~P. Gordon, R.~C.~C. Leite, R.~S. Moore, S.~P.~S. Porto, and J.~R. Whinnery,
  ``Long‐{Transient} {Effects} in {Lasers} with {Inserted} {Liquid}
  {Samples},'' {\em Journal of Applied Physics}, vol.~36, pp.~3--8, Jan. 1965.

\bibitem{lugiato_bistability_1988}
L.~A. Lugiato, C.~Oldano, C.~Fabre, E.~Giacobino, and R.~J. Horowicz,
  ``Bistability, self-pulsing and chaos in optical parametric oscillators,''
  {\em Il Nuovo Cimento D}, vol.~10, pp.~959--977, Aug. 1988.

\bibitem{de_valcarcel_transverse_1996}
G.~J. de~Valcárcel, K.~Staliunas, E.~Roldán, and V.~J. Sánchez-Morcillo,
  ``Transverse patterns in degenerate optical parametric oscillation and
  degenerate four-wave mixing,'' {\em Physical Review A}, vol.~54,
  pp.~1609--1624, Aug. 1996.

\bibitem{haragus_local_2011}
M.~Haragus and G.~Iooss, {\em Local {Bifurcations}, {Center} {Manifolds}, and
  {Normal} {Forms} in {Infinite}-{Dimensional} {Dynamical} {Systems}}.
\newblock Universitext, London: Springer-Verlag, 2011.

\bibitem{turing_alan_mathison_chemical_1952}
{Turing Alan Mathison}, ``The chemical basis of morphogenesis,'' {\em
  Philosophical Transactions of the Royal Society of London. Series B,
  Biological Sciences}, vol.~237, pp.~37--72, Aug. 1952.

\bibitem{tlidi_spatiotemporal_1997}
M.~Tlidi, P.~Mandel, and M.~Haelterman, ``Spatiotemporal patterns and localized
  structures in nonlinear optics,'' {\em Physical Review E}, vol.~56,
  pp.~6524--6530, Dec. 1997.

\bibitem{tlidi_robust_1998}
M.~Tlidi and M.~Haelterman, ``Robust {Hopf}-{Turing} mixed-mode in optical
  frequency conversion systems,'' {\em Physics Letters A}, vol.~239,
  pp.~59--64, Feb. 1998.

\bibitem{champneys_homoclinic_1998}
A.~R. Champneys, ``Homoclinic orbits in reversible systems and their
  applications in mechanics, fluids and optics,'' {\em Physica D: Nonlinear
  Phenomena}, vol.~112, pp.~158--186, Jan. 1998.

\bibitem{chomaz_absolute_1992}
J.~M. Chomaz, ``Absolute and convective instabilities in nonlinear systems,''
  {\em Physical Review Letters}, vol.~69, pp.~1931--1934, Sept. 1992.

\bibitem{allen_microscopic_1979}
S.~M. Allen and J.~W. Cahn, ``A microscopic theory for antiphase boundary
  motion and its application to antiphase domain coarsening,'' {\em Acta
  Metallurgica}, vol.~27, pp.~1085--1095, June 1979.

\bibitem{coullet_breaking_1990}
P.~Coullet, J.~Lega, B.~Houchmandzadeh, and J.~Lajzerowicz, ``Breaking
  chirality in nonequilibrium systems,'' {\em Physical Review Letters},
  vol.~65, pp.~1352--1355, Sept. 1990.

\bibitem{gomila_theory_2015}
D.~Gomila, P.~Colet, and D.~Walgraef, ``Theory for the {Spatiotemporal}
  {Dynamics} of {Domain} {Walls} close to a {Nonequilibrium} {Ising}-{Bloch}
  {Transition},'' {\em Physical Review Letters}, vol.~114, p.~084101, Feb.
  2015.

\bibitem{morgan_swifthohenberg_2014}
D.~Morgan and J.~H.~P. Dawes, ``The {Swift}–{Hohenberg} equation with a
  nonlocal nonlinearity,'' {\em Physica D: Nonlinear Phenomena}, vol.~270,
  pp.~60--80, Mar. 2014.

\bibitem{allgower_numerical_1990}
E.~L. Allgower and K.~Georg, {\em Numerical {Continuation} {Methods}: {An}
  {Introduction}}.
\newblock Springer {Series} in {Computational} {Mathematics}, Berlin
  Heidelberg: Springer-Verlag, 1990.

\bibitem{knobloch_homoclinic_2005}
J.~Knobloch and T.~Wagenknecht, ``Homoclinic snaking near a heteroclinic cycle
  in reversible systems,'' {\em Physica D: Nonlinear Phenomena}, vol.~206,
  pp.~82--93, June 2005.

\bibitem{ma_defect-mediated_2010}
Y.~P. Ma, J.~Burke, and E.~Knobloch, ``Defect-mediated snaking: {A} new growth
  mechanism for localized structures,'' {\em Physica D: Nonlinear Phenomena},
  vol.~239, pp.~1867--1883, Oct. 2010.

\bibitem{parra-rivas_dark_2016}
P.~Parra-Rivas, E.~Knobloch, D.~Gomila, and L.~Gelens, ``Dark solitons in the
  {Lugiato}-{Lefever} equation with normal dispersion,'' {\em Physical Review
  A}, vol.~93, p.~063839, June 2016.

\bibitem{godey_bifurcation_2017}
C.~Godey, ``A bifurcation analysis for the {Lugiato}-{Lefever} equation,'' {\em
  The European Physical Journal D}, vol.~71, p.~131, May 2017.

\bibitem{parra-rivas_bifurcation_2018}
P.~Parra-Rivas, D.~Gomila, L.~Gelens, and E.~Knobloch, ``Bifurcation structure
  of localized states in the {Lugiato}-{Lefever} equation with anomalous
  dispersion,'' {\em Physical Review E}, vol.~97, p.~042204, Apr. 2018.

\bibitem{kozyreff_asymptotics_2006}
G.~Kozyreff and S.~J. Chapman, ``Asymptotics of {Large} {Bound} {States} of
  {Localized} {Structures},'' {\em Physical Review Letters}, vol.~97,
  p.~044502, July 2006.

\bibitem{woods_heteroclinic_1999}
P.~D. Woods and A.~R. Champneys, ``Heteroclinic tangles and homoclinic snaking
  in the unfolding of a degenerate reversible {Hamiltonian}–{Hopf}
  bifurcation,'' {\em Physica D: Nonlinear Phenomena}, vol.~129, pp.~147--170,
  May 1999.

\bibitem{burke_snakes_2007}
J.~Burke and E.~Knobloch, ``Snakes and ladders: {Localized} states in the
  {Swift}–{Hohenberg} equation,'' {\em Physics Letters A}, vol.~360,
  pp.~681--688, Jan. 2007.

\bibitem{parra-rivas_coexistence_2017}
P.~Parra-Rivas, D.~Gomila, and L.~Gelens, ``Coexistence of stable dark- and
  bright-soliton {Kerr} combs in normal-dispersion resonators,'' {\em Physical
  Review A}, vol.~95, p.~053863, May 2017.

\bibitem{kuehn_validity_2018}
C.~Kuehn and S.~Throm, ``Validity of amplitude equations for nonlocal
  nonlinearities,'' {\em Journal of Mathematical Physics}, vol.~59, p.~071510,
  July 2018.

\end{thebibliography}
\end{document}